\documentclass[12pt,notitlepage]{iopart}

\usepackage[noadjust]{cite}
\usepackage{graphicx,color}
\usepackage{amssymb}

\newcommand{\eqref}[1]{(\ref{#1})}

\begin{document}

\title{Spin-gap spectroscopy in a bosonic flux ladder}

\author{Marcello~Calvanese~Strinati$^{1,2}$, Fabrice~Gerbier$^3$ and Leonardo~Mazza$^2$}
\address{$^1$NEST, Scuola Normale Superiore \& Istituto Nanoscienze-CNR, I-56126 Pisa, Italy}
\address{$^2$D\'epartement de Physique, Ecole Normale Sup\'erieure / PSL Research University,
CNRS, 24 rue Lhomond, F-75005 Paris, France}
\address{$^3$Laboratoire Kastler Brosser, Coll\`ege de France, CNRS, ENS-PSL Research University,
UPMC-Sorbonne Universit\`es, 11 place Marcelin Berthelot, {F-}75005 Paris, France}


\begin{abstract}
Ultracold bosonic atoms trapped in a two-leg ladder pierced by a magnetic field provide a minimal and quasi-one-dimensional instance to study the interplay between orbital magnetism and {interactions}. Using time-dependent matrix-product-states simulations, we investigate the properties of the so-called ``Meissner'' and ``vortex'' phases which appear in such system, focusing on experimentally accessible observables. We discuss how to experimentally monitor the phase transition, and show that the response to a modulation of the density imbalance between the two legs of the ladder is qualitatively different in the two phases. We argue that this technique can be used as a tool for many-body spectroscopy, allowing to quantitatively measure the spin gap in the Meissner phase. We finally discuss its experimental implementation.
\end{abstract}


\section{Introduction}

Orbital magnetism (OM) encompasses a host of phenomena that arise in systems of charged particles subject to an applied magnetic field. {Because} the Bohr-van Leeuwen theorem forbids its appearance in an ensemble of classical particles~\cite{borhcollectedworks,jphysradium2361}, {OM is} a trademark of quantum mechanics since its early days{. I}n the case of electrons in solids, for instance, OM effects include Landau diamagnetism~\cite{landau2013statistical}, and the integer and fractional quantum Hall effects~\cite{PhysRevLett.45.494,PhysRevLett.48.1559}.

Flux ladders (FL) composed of two (or more) coupled one-dimensional subparts with a magnetic field perpendicular to the ladder plane are among the simplest setups where OM can appear. FL are quasi-one-dimensional, and thus still amenable to an efficient theoretical treatment in the presence of interactions, either using bosonization~\cite{giamarchi2003quantum} or numerical methods based on matrix-product states (MPS)~\cite{RevModPhys.77.259,Schollwock201196}. {Establishing the connection with two-dimensional physics for studying FL is one of the major motivations in this research field.}

Bosonic two-leg FLs have been particularly studied, in part due to the simplicity of the model, and in part because of the recent experimental realization with ultracold atoms in suitably designed optical lattices \cite{NatPhys10.588.14}. Using the bosonization technique, the pioneering work of~\cite{PhysRevB.64.144515} predicts the appearance of vortex (V) and Meissner (M) phases paralleling the phenomenology of superconductors. A V phase is characterized by non-vanishing inter-leg (``transverse'') current, and a M phase by vanishing transverse current. For strong interactions and commensurate densities, a phase transition between a Mott-insulator (MI) and a superfluid (SF) also appears~\cite{PhysRevB.63.180508}. 
According to these field-theory treatments of the low-energy part of the model, two-leg ladders feature generally two excitation branches, related to ``charge'' (or ``density'') degrees of freedom on the one hand, and to ``spin'' degrees of freedom on the other~\cite{PhysRevB.64.144515,PhysRevB.63.180508}. The MI phases then correspond to the opening of a charge gap, and the M phases to the opening of a spin gap. All the four situations obtained by combining these two classifications -- V-SF, M-SF, V-MI and M-MI -- are possible. 
Numerical studies of microscopic models of interacting bosonic FLs { have confirmed the existence of these four phases and more,} revealing an extraordinarily rich phenomenology~\cite{PhysRevB.72.104521,PhysRevB.73.100502, PhysRevA.85.041602, PhysRevLett.111.150601, PhysRevB.87.174501, PhysRevA.89.063617, 1367-2630-16-7-073005, PhysRevB.91.054520, DiDio2015, PhysRevB.91.140406, PhysRevA.92.013625, PhysRevB.92.060506, PhysRevB.92.115446, 1367-2630-17-9-092001, PhysRevB.92.115120, PhysRevA.92.053623, PhysRevLett.115.190402, PhysRevA.93.053629, PhysRevA.94.023630, PhysRevA.94.063628,1367-2630-18-5-055017,PhysRevA.94.063632}. For instance, it has been proposed { recently} that precursors of the physics of the fractional quantum Hall effect, and in particular of Laughlin wave functions, might appear in experimentally-relevant bosonic FL~\cite{PhysRevB.91.054520,PhysRevB.92.115446,PhysRevX.7.021033,PhysRevB.96.014524}.

Experimentally, the realization of bosonic FL belongs to a more general effort to realize effective gauge potentials coupling to ultracold atoms in spite of their electrical neutrality \cite{RevModPhys.83.1523,0034-4885-77-12-126401}. The experiment of~\cite{NatPhys10.588.14} creates a one-dimensional array of isolated ladders with a total flux per plaquette $\Phi = \pi/2$ induced by combining laser-assisted hopping with a periodic spatial modulation of the lattice. In this experiment, each site of the ladder is in reality a one-dimensional bosonic gas with many atoms, with the result that the interaction energy per atom was very weak compared to inter- and intra-leg tunneling energies. Recently, the role of interactions in bosonic FL was {experimentally} investigated for two particles \cite{nature22811}.

The experiment of~\cite{science1514} exploits the concept of ``synthetic dimension''. Each leg of the ladder can be represented by internal (spin) states of the atom, and the magnetic flux is due to Raman transitions coupling the internal states. The idea of synthetic dimension has been recently generalized to momentum space lattices \cite{Fangzhao2017a}. Importantly, in the synthetic dimension approach, the two legs are not separated in space, but fully overlapping. As a result, interactions  are short-ranged in real space, but have almost infinite range along the synthetic dimension. This makes interacting models using the synthetic dimension approach quite different from models with short-range interactions~\cite{PhysRevLett.112.043001,1367-2630-18-3-035010}.

{Fermionic flux ladders can also be explored experimentally with ultracold atoms using similar approaches as in the bosonic case~\cite{Zhang1467,NatPhysPagano,science1510,PhysRevLett.117.220401}. Theoretical studies have highlighted the presence of fractional charge excitations and predicted a host of novel phases of matter (such as charge-, bond- and density- waves or orbital antiferromagnets) leading to a more complex phenomenology than the V-M competition of the bosonic case~\cite{PhysRevB.71.161101,PhysRevB.73.195114,PhysRevB.76.195105}. Triggered by the interest in the quantum Hall effect, analogues of the chiral modes which characterize both integer and fractional phases have been discussed~\cite{1367-2630-17-10-105001,PhysRevB.92.115446,ncomms9134,1367-2630-18-3-035010,PhysRevB.92.245121,PhysRevA.93.013604,PhysRevA.95.063612,PhysRevX.7.021033,arXiv:1707.05715}.}

In this article, we propose an experimentally-feasible {method} to distinguish the M and V phases {in the bosonic FL} and to characterize their low-energy excitation spectrum. It is known that the M and V phases can be distinguished qualitatively by time-of-flight methods~\cite{RevModPhys.80.885,PhysRevLett.107.255301,NatPhys10.588.14,science1510,1367-2630-18-5-055017,PhysRevA.94.063628}. We show that they {also} respond differently to a periodic ``spin'' modulation, and we interpret our results as a measure of the spin gap in the M phase. We {support our claims by presenting} numerical simulations performed both in the dilute non-interacting limit and in the dense interacting case. This extends previous work studying dynamical protocols to probe bosonic or fermionic systems in one dimension~\cite{PhysRevA.74.041604,PhysRevLett.97.050402,PhysRevA.73.041608,PhysRevLett.97.260401,PhysRevB.77.245119,10.1088/0953-4075/46/8/085303}. Finally, we show how to adapt the proposal of~\cite{njp.12.033007}, initially designed to realize two-dimensional systems with an effective magnetic flux, to the realization of FL with strong {on-site} interactions. This scheme is well suited to the spectroscopic method probing the spin gap, although we note that the method can also be used in other implementations of bosonic FL.

The article is organized as follows. In Section~\ref{sec:themodel}, we introduce the model, and in Section~\ref{sec:momentumdistributionfunction} we briefly discuss some aspects of its phase diagram. In Section~\ref{sec:spectroscopy}, we present our theory for the spin-gap spectroscopy and the numerical simulations supporting our statements. In Section~\ref{sec:experimentalrealizability}, we discuss a possible experimental implementation of bosonic FL using state-dependent lattices and laser-induced tunneling, and discuss how the proposed measurement could be carried out. We finally draw our conclusions in Section~\ref{sec:conclusions}, {and provide some technical details in the appendices}.


\section{Model and notations}
\label{sec:themodel}

\begin{figure}[t]
	\centering
	\includegraphics[width=11cm]{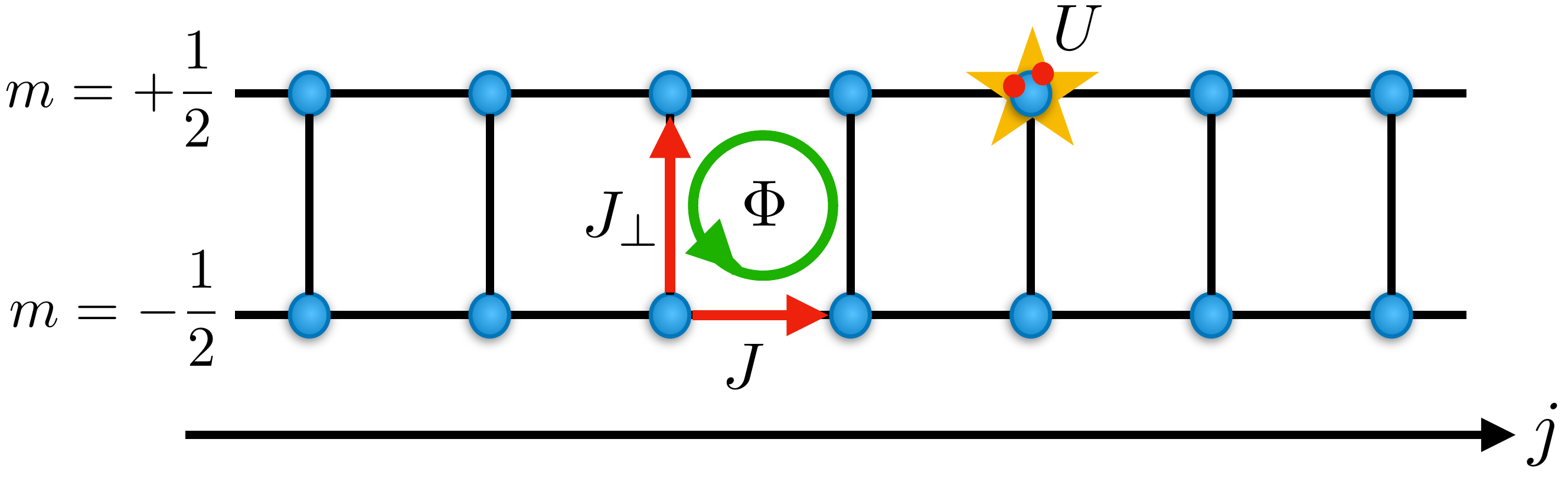}
	\caption{Schematic representation of the two-leg ladder. Here, $J$ is the tunneling amplitude between nearest-neighbor sites in the longitudinal direction $j$, $J_\perp$ is the tunneling amplitude in the transverse direction $m$, $\Phi$ is the gauge flux piercing each plaquette, and $U$ is the on-site interaction strength, taken equal for both legs.}
	\label{fig:setupandladderscheme}
\end{figure}

We consider a gas of interacting bosonic atoms loaded into an optical lattice at zero temperature. The system is a FL composed of two coupled one-dimensional systems immersed in a (possibly synthetic) magnetic field. A sketch of the ladder is shown in Figure~\ref{fig:setupandladderscheme}, where $j$ and $m$ identify the longitudinal and transverse directions of the ladder respectively. Such a system can be modeled by the following tight-binding Hamiltonian including interactions~\cite{PhysRevB.64.144515}:
\begin{eqnarray}
\hat H_0^{(\rm ex)}&=&-J\sum_{j=1}^{L-1}\,\sum_{m=\pm1/2}\left(\hat b^\dag_{j,m}\hat b_{j+1,m}+{\rm H.c.}\right)+J_\perp\sum_{j=1}^{L}\left(\hat b^\dag_{j,-\frac{1}{2}}\hat b_{j,+\frac{1}{2}}\,e^{-i\Phi j}+{\rm H.c.}\right)\nonumber\\
\nonumber\\
&&+\frac{U}{2}\sum_{j=1}^{L}\,\sum_{m=\pm1/2}\hat n_{j,m}\left(\hat n_{j,m}-1\right).
\label{eq:bosonictwolegladderhamiltonian}
\end{eqnarray}
Here, $\hat b_{j,m}$ ($\hat b^\dag_{j,m}$) annihilates (creates) a boson on site $j$ and on the leg $m$, $\hat n_{j,m}=\hat b^\dag_{j,m}\hat b_{j,m}$ is the local density operator on the leg $m$, $J$ and $J_\perp$ denote the tunneling amplitude between two nearest-neighbor (NN) sites in the longitudinal and transverse direction, respectively, and $\Phi$ is the magnetic flux per plaquette. The inter-particle interaction is taken into account by the Bose-Hubbard on-site interaction $U$, which we take equal for both legs. We denote by $L$ the total number of rungs of the ladder, and consider open boundary conditions (OBC). The total number of particles $N$ defines the particle density per rung $n$ through $n=N/L$.

In the Hamiltonian in Equation~\eqref{eq:bosonictwolegladderhamiltonian}, the gauge flux is set in such a way that the tunneling matrix elements on the transverse links of the ladder are complex, and the longitudinal ones are real. We will refer to this choice as the \emph{experimental gauge} (ex). It is convenient to make the Hamiltonian in Equation~\eqref{eq:bosonictwolegladderhamiltonian} translationally invariant, swapping the gauge flux to the longitudinal links, by using the unitary transformation $\hat d_{j,m}=e^{-i\Phi jm}\,\hat b_{j,m}$. The transformed Hamiltonian reads
\begin{eqnarray}
\hat H_0^{({\rm cm})}&=&-J\sum_{j}\,\sum_{m=\pm1/2}\left(\hat d^\dag_{j,m}\hat d_{j+1,m}\,e^{i\Phi m}+{\rm H.c.}\right)+J_\perp\sum_j\left(\hat d^\dag_{j,-\frac{1}{2}}\hat d_{j,+\frac{1}{2}}+{\rm H.c.}\right)\nonumber\\
\nonumber\\
&&+\frac{U}{2}\sum_j\sum_{m=\pm1/2}\hat n_{j,m}\left(\hat n_{j,m}-1\right) \,\, ,
\label{eq:bosonictwolegladderhamiltoniancmgauge}
\end{eqnarray}
where $\hat n_{j,m}=\hat b^\dag_{j,m}\hat b_{j,m}=\hat d_{j,m}^\dag\hat d_{j,m}$. The choice of the gauge as in Equation~\eqref{eq:bosonictwolegladderhamiltoniancmgauge} will be referred to as the \emph{condensed-matter gauge} (cm). In the following, if not explicit, we use $J$ as reference energy scale.

For non-interacting bosons ($U=0$), the Hamiltonian in Equation~\eqref{eq:bosonictwolegladderhamiltoniancmgauge} can be diagonalized in momentum space by introducing the operators $\hat d_{k,m}=L^{-1/2}\sum_je^{ikj}\,\hat d_{j,m}$. The two energy bands are given by $E_\pm(k)=-2J\cos(k)\cos(\Phi/2)\pm\sqrt{4J^2\sin^2(k)\sin^2(\Phi/2)+J_\perp^2}$. The structure of the lower band $E_-(k)$ changes with $J_\perp$ or $\Phi$. When $J_\perp$ exceeds a critical value $J_{\perp,c}=2J\sin(\Phi/2)\tan(\Phi/2)$, the lower energy band has one minimum at $k=0$. When $J_\perp < J_{\perp,c}$, the lower band features two symmetric minima at $k=\pm k_M(\Phi,J_\perp)$. In the former case, the system is in the \emph{Meissner phase} (M), whereas it is in the \emph{vortex phase} (V) in the latter. By tuning $J_\perp$ and/or $\Phi$, the system can undergo the M-V phase transition~\cite{1367-2630-16-7-073005}. This transition persists for non-zero repulsive interactions, but the critical value $J_{\perp,c}$ (that depends on $U,n,\Phi$ in general) can be strongly modified by interactions~\cite{PhysRevLett.115.190402,PhysRevA.94.063628}. 

\section{Momentum distribution functions and phase diagram of interacting bosonic flux ladders}
\label{sec:momentumdistributionfunction}

We numerically study the properties of bosonic FL with repulsive interactions ($U>0$) using a MPS-based algorithm~\cite{Schollwock201196}. The ground state (GS) of the system is found after a local variational search in the MPS space. At finite $U$, we keep $d_{\rm loc}=3$ states for the local Hilbert space (see \ref{sec:appendix1} for details and a critical discussion).

According to bosonization, the M phase is distinguished from the V phase by {the presence of a gap appearing in the spin sector of the low-energy theory}~\cite{PhysRevB.64.144515,PhysRevB.91.140406} (hereafter denoted as ``spin gap''). As a consequence, the two phases differ also in the so-called \emph{central charge} $c$ {that, in this context,} roughly speaking gives {half} the number of gapless modes~\cite{giamarchi2003quantum}. When the particle density is less than unity, $n<1$ (which is the situation that will be studied in this article), the charge sector is always gapless (no MI phase). The spin sector is gapped in the M phase (thus $c=1$ if $n<1$), and gapless in the V phase (thus $c=2$ if $n<1$). Monitoring the change of $c$ with variations of parameters $J_\perp,\Phi,n$ allows one to track the M-V phase transition (see \ref{sec:appendix1}). MPS methods are well-suited to extract the entanglement entropy from which the central charge is deduced~\cite{1742-5468-2004-06-P06002}.

{Another possibility would be the direct numerical computation of the spin gap that distinguishes the two phases. Such measurement is typically performed in ladder or more general models with two decoupled species, where the number of particles for each species is a conserved quantity~\cite{PhysRevB.57.10324,PhysRevB.83.205113}. In our situation, however, the spin gap can not be accessed directly: only the total number of particles is a conserved quantity when  $J_\perp$ and $\Phi$ are both non-zero. As a result, there is no quantum number associated with the spin sector (outside of the low-energy sector). This makes the computation of the spin gap unfeasible in practice. We propose in the next section a spectroscopic method that can be used to estimate the spin gap.}

We begin by reviewing a method to study the phase diagram~\cite{PhysRevLett.115.190402,PhysRevA.94.063628,PhysRevB.92.060506,1367-2630-18-5-055017,DiDio2015}, which can be easily implemented in experiments~\cite{NatPhys10.588.14}. We focus on the momentum distribution functions (MDF), both leg-resolved and total. Time-of-flight measurements readily give access to the total MDF; in some experimental schemes, such as the one discussed in Section~\ref{sec:experimentalrealizability}, is even possible to measure it only for a specific leg. The leg-resolved MDF in the experimental gauge is defined as
\begin{equation}
n^{({\rm ex})}_m(k)=\left\langle\hat b^\dag_{k,m}\hat b_{k,m}\right\rangle=\frac{1}{L}\sum_{j,h=1}^{L}e^{-ik(j-h)}\,\left\langle\hat b^\dag_{j,m}\hat b_{h,m}\right\rangle \,\, ,
\label{eq:momentumsidtributionfunctioncagauge}
\end{equation}
where the expectation value is computed over the GS of the Hamiltonian in Equation~\eqref{eq:bosonictwolegladderhamiltonian}. Since the MDF is periodic with period $2\pi$, we restrict the momentum variable to $k\in[-\pi:\pi)$. By using the unitary transformation introduced before, the MDF in the experimental and in the condensed-matter gauge are simply related by a momentum shift, i.e., $n^{({\rm cm})}_m(k)=n^{({\rm ex})}_m(k-m\Phi)$. The total MDFs are accordingly $n^{({\rm ex})}(k)=\sum_{m=\pm1/2}n^{({\rm ex})}_m(k)$ and $n^{({\rm cm})}(k)=\sum_{m=\pm1/2}n^{(\rm cm)}_m(k)$.

\begin{figure}[t]
\centering
\includegraphics[width=7.7cm]{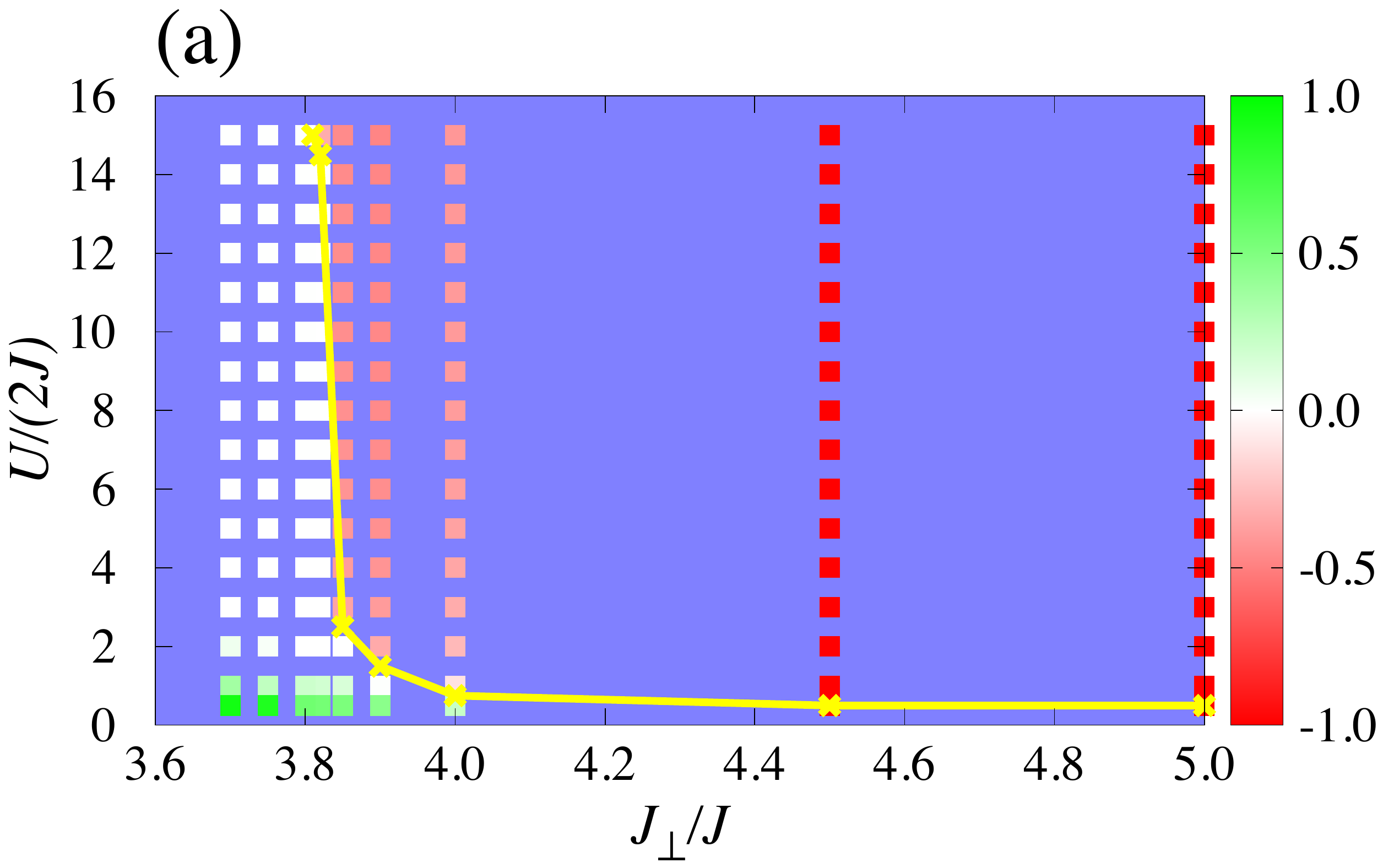}
\includegraphics[width=7.7cm]{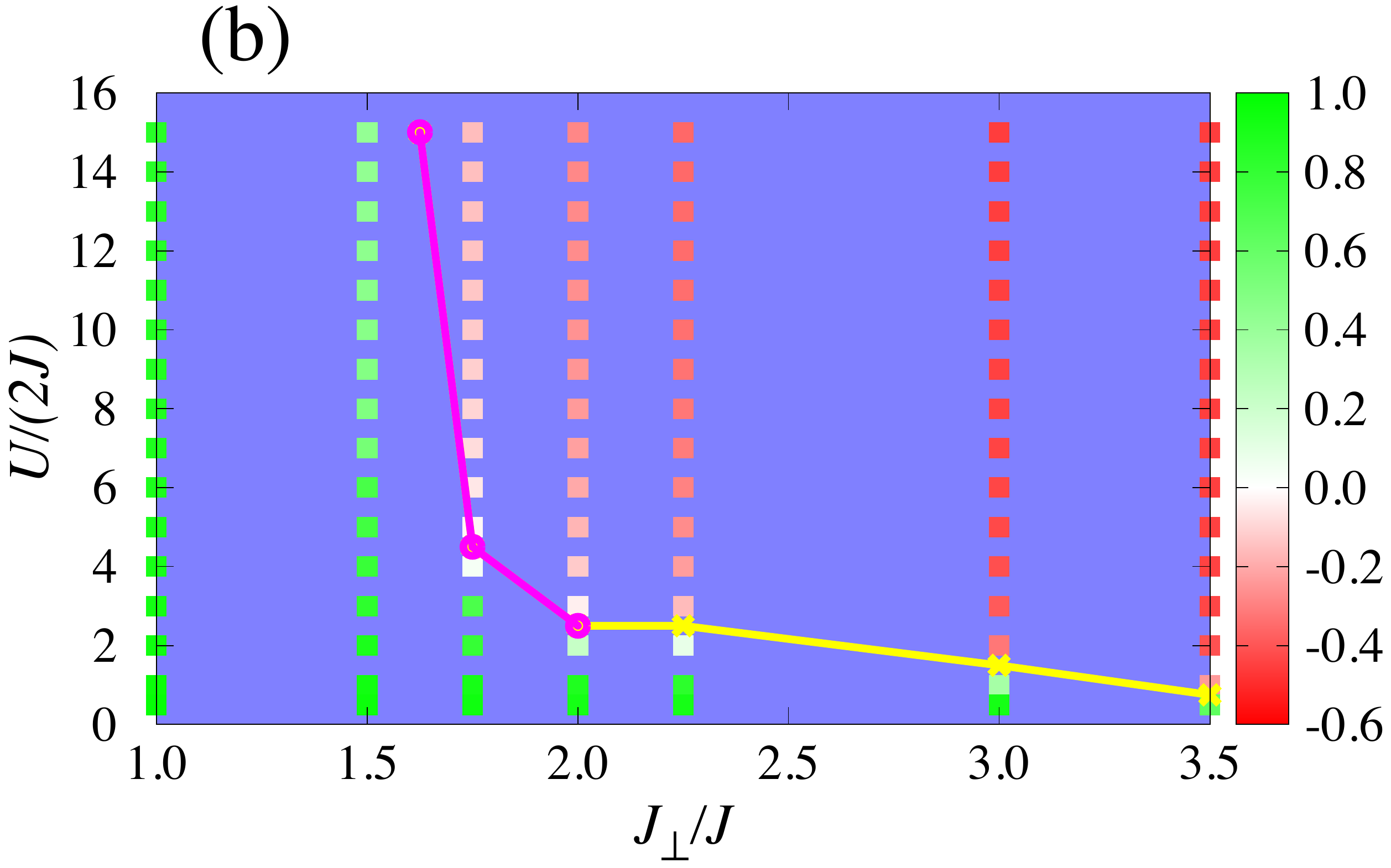}
\includegraphics[width=7.7cm]{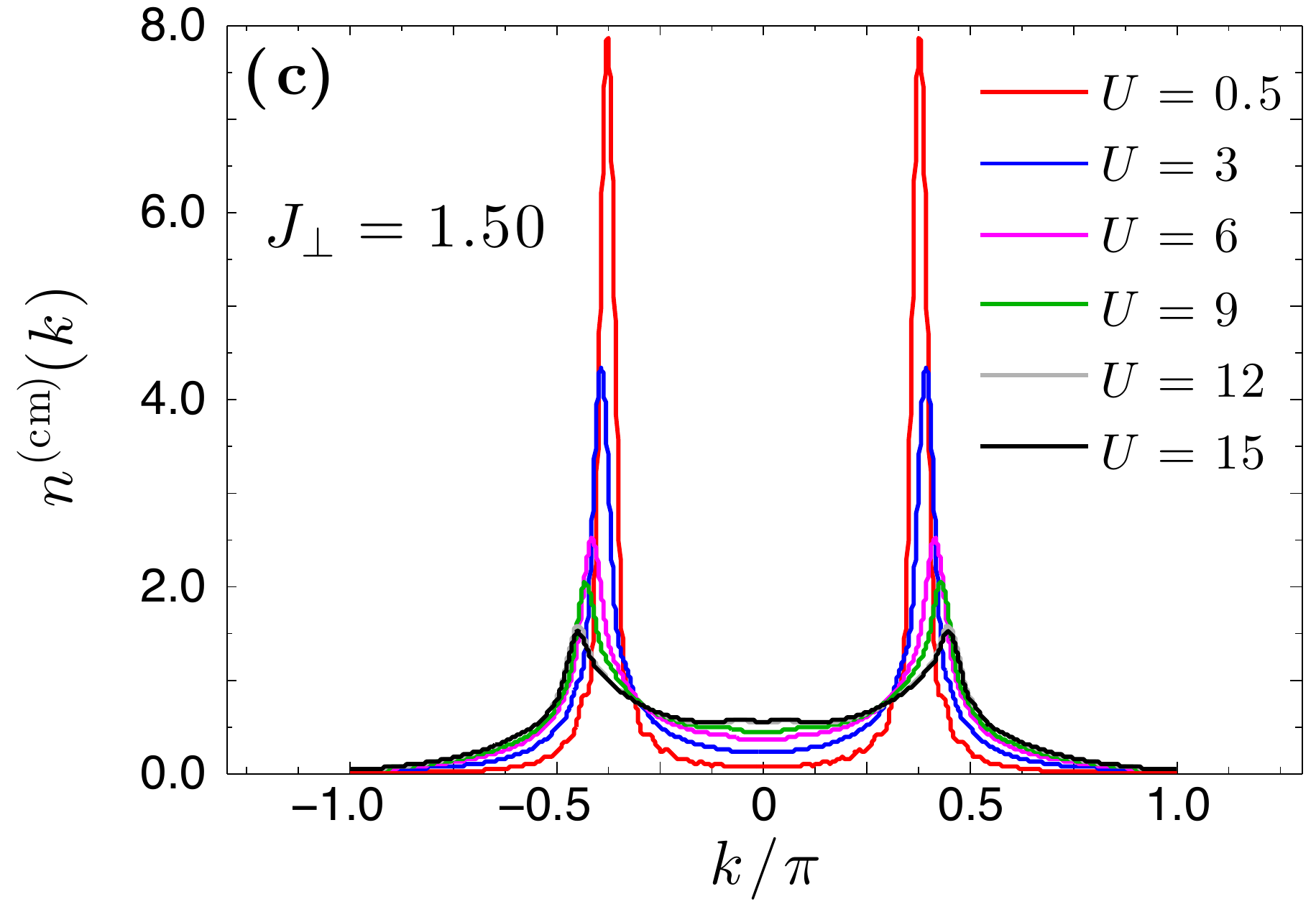}
\includegraphics[width=7.7cm]{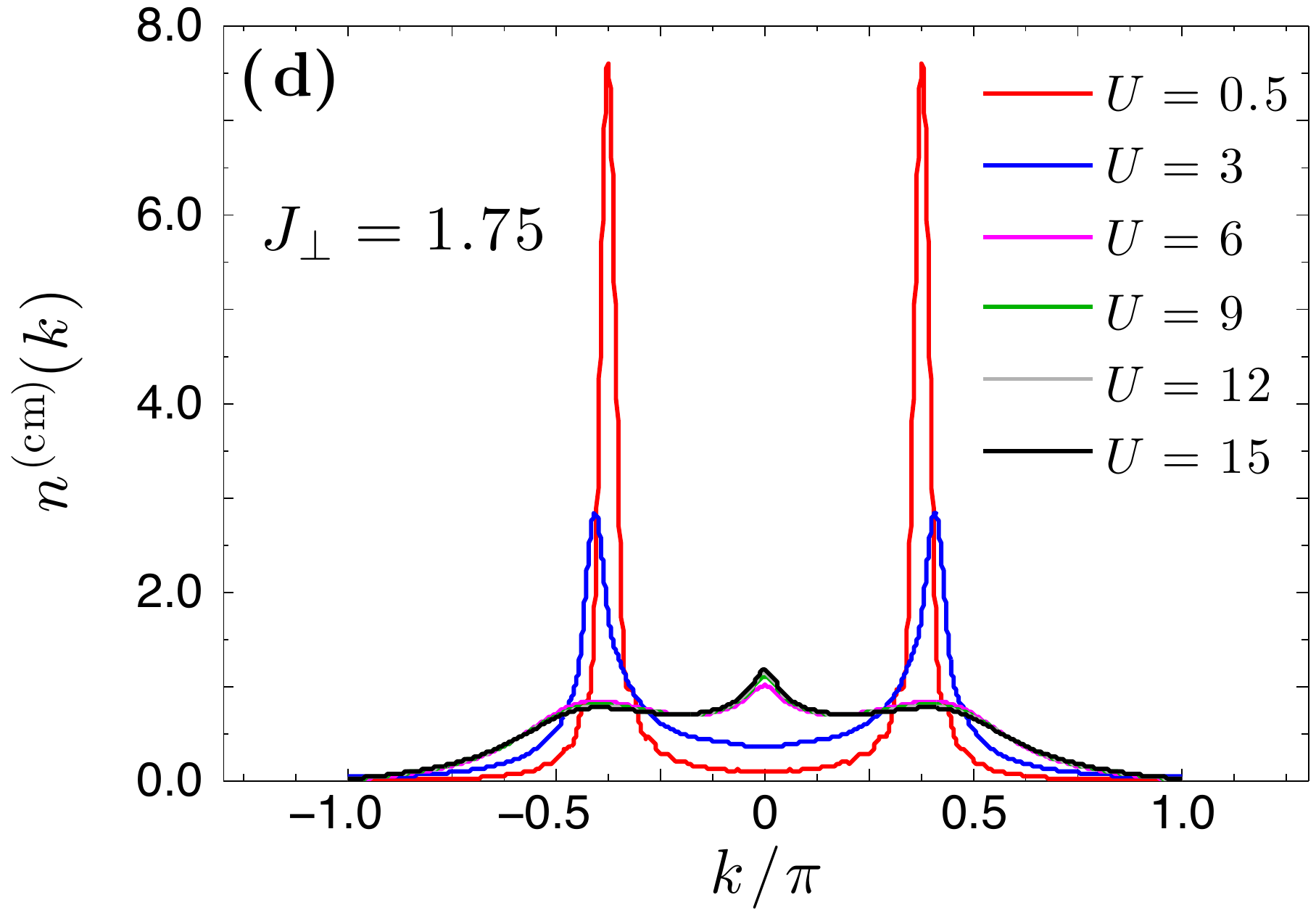}
\caption{Phase diagram in the $U/(2J)$ vs. $J_\perp/J$ plane. \textbf{(a)} We use ${n=1/4}$, $L=72$ and $\Phi=0.8\,\pi$. The points on the critical line (yellow line) are found by imposing $\delta n=0$. The green points identify the V phase, whereas red points identify the M phase. \textbf{(b)} Same analysis as in panel \textbf{(a)} but using ${n=1/2}$ and $L=24$. In the latter case, we have to use smaller values of $L$ because of the higher numerical complexity. The critical line is shifted towards smaller values of $J_\perp$. To increase the numerical accuracy, we compute the critical line using $L=48$ (magenta line), which overlaps with the one obtained using $L=24$ (yellow line). \textbf{(c)}-\textbf{(d)} Data for $n^{({\rm cm})}(k)$ for the phase diagram in panel \textbf{(b)}, using $L=48$, for \textbf{(c)} $J_\perp/J=1.50$, \textbf{(d)} $J_\perp/J=1.75$}
\label{fig:meissnerandvortexphases}
\end{figure}

In the condensed-matter gauge, the MDF displays one peak centered at $k=0$ for the M phase, and two symmetric peaks at $k=\pm k_M$ for the V phase, reminiscent of the single or double minimum of the lower energy band when $U=0$~\cite{1367-2630-18-5-055017}. We report in Figure~\ref{fig:meissnerandvortexphases} the MDF for $J_\perp/J=1.50$ (panel \textbf{(c)}) and  $J_\perp/J=1.75$ (panel \textbf{(d)}) for several values of $U$ and ${n = 1/2}$. For sufficiently low values of $J_\perp$, the two-peak structure of the MDF is observed for all $U$. For large enough $J_\perp$, the V-M phase transition occurs when $U$ is increased beyond a critical value; in this case, we see the emergence of a third peak at $k=0$, which eventually dominates the MDF when one enters the M phase. 

To go beyond these qualitative features and to quantitatively distinguish M and V phases, we define the imbalance ratio (IR)
\begin{equation}
\delta n(\Phi,J_\perp,U,n):=\frac{n^{({\rm cm})}(k_M)-n^{({\rm cm})}(0)}{n^{({\rm cm})}(k_M)+n^{({\rm cm})}(0)} \,\, .
\label{eq:quantityforphasediagram}
\end{equation}
The IR takes the values $0<\delta n<1$ in the V phase and $-1<\delta n<0$ in the M phase. We propose to find the transition points by imposing the condition $\delta n=0$. The IR provides a simple and experimentally accessible observable to distinguish V and M phases, although it is not an order parameter in the sense of Landau theory. A more rigorous numerical characterization of the two phases is provided in \ref{sec:appendix1}, where we show that, for ${n=1/2}$, the transition point identified by $\delta n=0$ is very close to the point where the central charge introduced earlier changes from $c=2$ to $c=1$~\cite{PhysRevB.91.140406}. By monitoring the variations of the IR with a control parameter, for instance $J_\perp$, we can obtain a qualitative phase diagram for the Hamiltonian in Equation~\eqref{eq:bosonictwolegladderhamiltonian}, and analyze how the presence of interactions affects the critical point at which the V-M phase transition occurs. A similar analysis was discussed in~\cite{PhysRevA.94.063628}.

The phase diagram in the $U$-$J_\perp$ plane for a fixed flux per plaquette of $\Phi=0.8\,\pi$ is shown in Figure~\ref{fig:meissnerandvortexphases}\textbf{(a)} for ${n=1/4}$  and \textbf{(b)} for ${n=1/2}$. Red points correspond to $\delta n<0$ (M phase), green points correspond to $\delta n>0$ (V phase), and the yellow line represents the critical line separating the two phases. We first focus on the case with ${n=1/4}$ (Figure~\ref{fig:meissnerandvortexphases}\textbf{(a)}). Repulsive interactions $U>0$ shift the critical value of $J_\perp$ with respect to the non-interacting case $J_{\perp,c}(U=0)\simeq5.9\,J$ \cite{1367-2630-16-7-073005}. We find that $J_{\perp,c}(U,n)$ is a monotonous and decreasing function of $U$, with $J_{\perp,c}(\infty,n)\simeq3.8\,J$ for {hard-core} bosons ($U \rightarrow \infty$) and $\Phi=0.8\pi$. For a larger particle density, the shift of $J_{\perp,c}(U,n)$ is expected to be enhanced further with respect to the ${n=1/4}$ case. The numerical simulation confirms this expectation, as we show in Figure~\ref{fig:meissnerandvortexphases}\textbf{(b)} for ${n=1/2}$.


\section{Spin gap spectroscopy} \label{sec:spectroscopy}

In the previous Section, we characterized the M and V phases by looking at the MDF. In this Section, we study the response of the bosonic ladder to a periodic imbalance of the particle number on the two legs, and we show that the system displays different responses in the M and V phases. We interpret our method as a spectroscopic tool that detects and measures the presence of the spin gap in the M phase predicted by bosonization.

\subsection{Model and observables}
\label{sec:modelandobservables}
We consider the Hamiltonian $\hat H_0^{({\rm ex})}$ in Equation~\eqref{eq:bosonictwolegladderhamiltonian}, and add a time-periodic perturbation $\hat V=F(t)\,\hat N_{s}$ proportional to the difference of populations between the two legs (hereafter denoted as \textit{spin imbalance}),
\begin{eqnarray}
\hat N_s=\hat N_{+\frac{1}{2}}-\hat N_{-\frac{1}{2}},
\label{eq:Ns}
\end{eqnarray}
with $\hat N_m=\sum_j\hat n_{j,m}$ the particle number per leg, with $F(t)=\delta_1\,\sin(\omega t)\,\Theta(t)$, and with $\Theta(t)$ the unit step function. Here, we denote by $\delta_1$ and $\omega$ the amplitude and frequency of the modulation, respectively. The total Hamiltonian is thus
\begin{eqnarray}
\hat H^{({\rm ex})}(t)=\hat H_0^{({\rm ex})}+\hat F(t)\,\hat N_s \,\, .
\label{eq:hamiltonianintheexperimentalgaugewiththeperiodicperturbation}
\end{eqnarray}
In what follows, to ease the notation, the superscripts denoting the experimental gauge in the Hamiltonian in Equation~\eqref{eq:hamiltonianintheexperimentalgaugewiththeperiodicperturbation} will be omitted. 

We consider the time evolution of the mean energy, $E(t)=\langle\Psi(t)|\hat H(t)|\Psi(t)\rangle$, where $|\Psi(t)\rangle$ is the time-evolved state starting from the GS of the bosonic ladder. We define the \emph{energy absorption rate} (EAR) as
\begin{eqnarray}
\label{eq:derivativeofthenergyintime}
\dot\varepsilon(\omega)=\lim_{T\rightarrow\infty}\frac{1}{T}\int_{0}^{T}dt\,\frac{\partial E}{\partial t} =
\lim_{T\rightarrow\infty}\frac{1}{T}\int_{0}^{T}dt\,\frac{\partial F}{\partial t}\langle\Psi(t)|\hat N_s|\Psi(t)\rangle \,\, .
\label{eq:energyabrosptionrate}
\end{eqnarray}
Within linear response theory, the EAR {per unit frequency} probes the imaginary part of the response function, i.e. $\dot\varepsilon(\omega)/\omega\propto{\rm Im}[\chi_{N_s-N_s}(\omega)]$, where $\chi_{N_s-N_s}(\omega)=\int_{0}^{\infty}dt\,e^{i\omega t}\,\chi_{N_s-N_s}(t)$ and where
\begin{eqnarray}
\chi_{N_s-N_s}(t)=-\frac{i}{\hbar}\,\langle\Psi_0|[\hat N_s(t),\hat N_s(0)]|\Psi_0\rangle
\label{eq:responsefunction}
\end{eqnarray}
is the response function in real time. Here $\hat N_s(t)=e^{i\hat H_0t/\hbar}\,\hat N_s\, e^{-i\hat H_0t/\hbar}$ is the number imbalance expressed in the interaction picture with respect to $\hat{H}_0$.  Notice that, by means of Equation~\eqref{eq:derivativeofthenergyintime}, the EAR can be experimentally accessed by measuring the total spin imbalance in time $\langle\Psi(t)|\hat N_s|\Psi(t)\rangle$ (see Section~\ref{sec:experimentalrealizability}).

If we denote by $\Delta E_s$ the value of the spin gap, a spectroscopic method that identifies it should consist of a periodic modulation of the system that is sensitive to its presence, so that the system does not absorb energy as long as $\hbar \omega<\Delta E_s$, and energy absorption can occur only for $\hbar \omega>\Delta E_s$.
We thus expect ${\rm Im}[\chi_{N_s-N_s}(\omega)]=0$ if $\hbar\omega<\Delta E_s$ and ${\rm Im}[\chi_{N_s-N_s}(\omega)]>0$ otherwise.

To compute the response in time to the modulation in Equation~\eqref{eq:hamiltonianintheexperimentalgaugewiththeperiodicperturbation}, we first compute the GS by means of the variational MPS-based algorithm discussed in Section~\ref{sec:themodel}. The time-evolved state, $|\Psi(t)\rangle$ is computed by using the time-evolving-block-decimation (TEBD) algorithm~\cite{Schollwock201196,PhysRevLett.91.147902,PhysRevLett.93.040502} with a fourth-order Trotter expansion~\cite{suzuki.prog.theor.phys.56.1454,suzuki.j.math.phys.32.400} with time step $dt$ (during the time evolution, we fix the maximum bond link $D_{{\rm max},t}$ used to describe the MPS state at time $t$).

\subsection{Results for dilute gases}
\label{sec:resultsforthedilutegascase}

\begin{figure}[t]
\centering
\includegraphics[width=7.7cm]{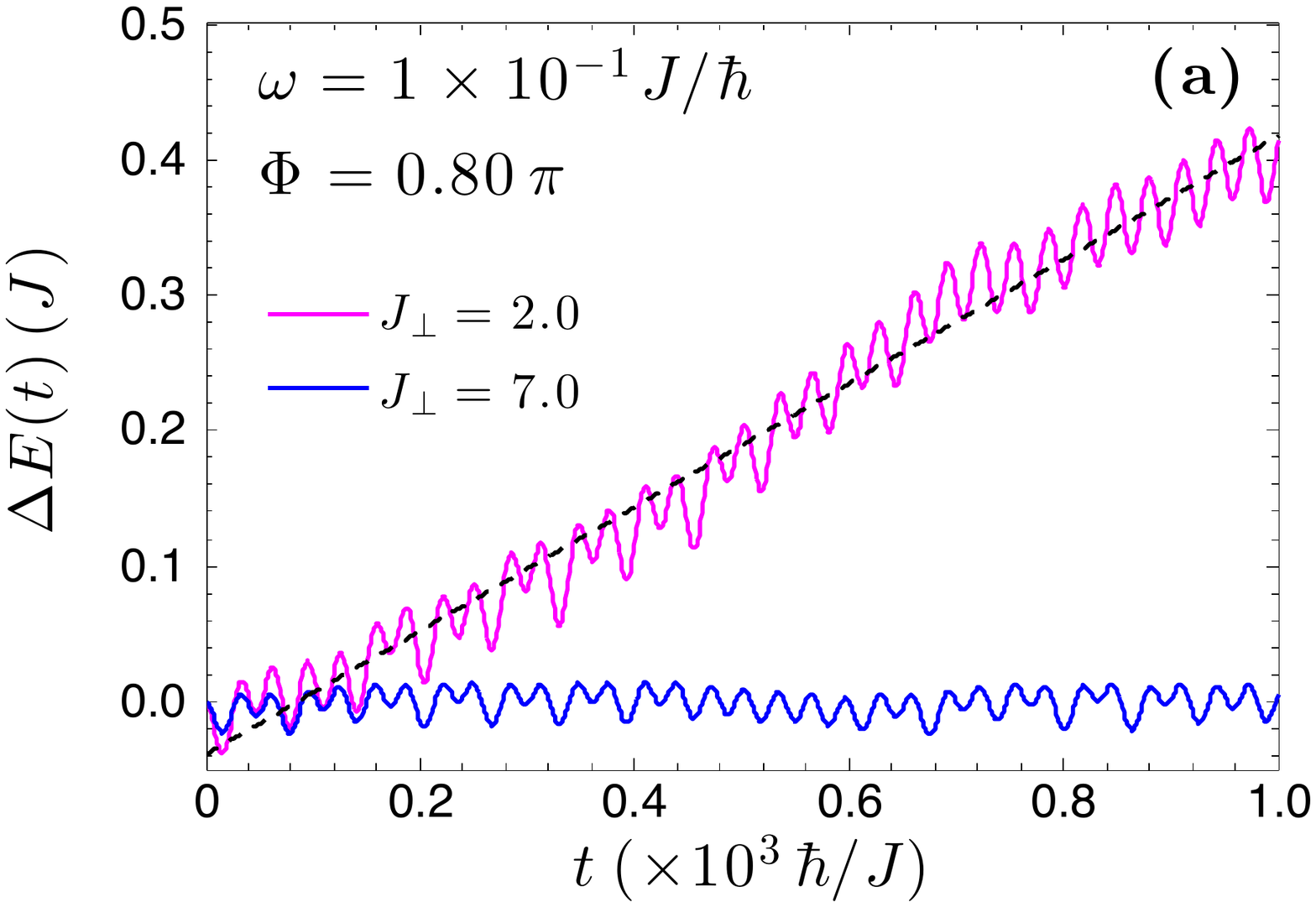}
\includegraphics[width=7.7cm]{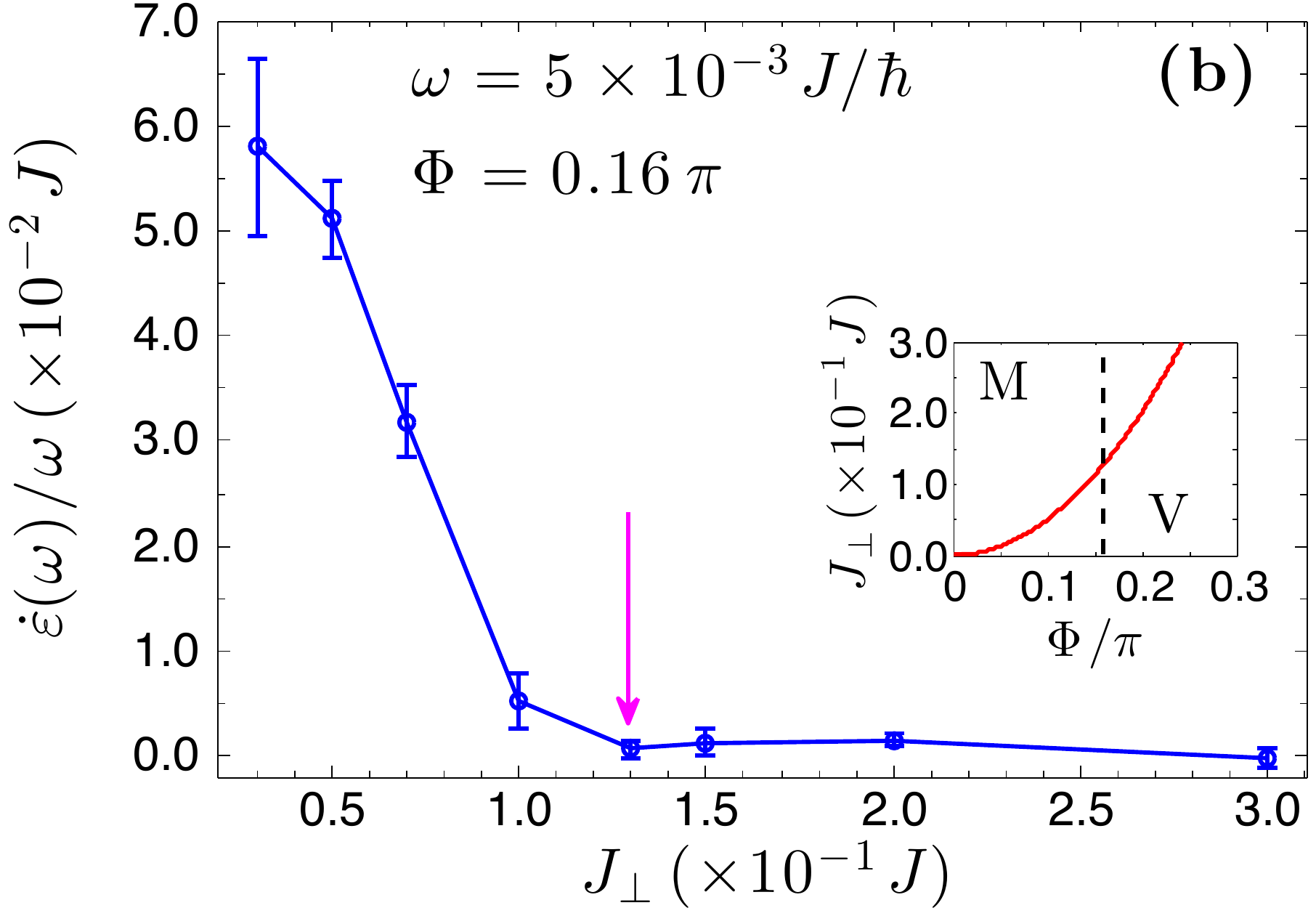}
\includegraphics[width=7.7cm]{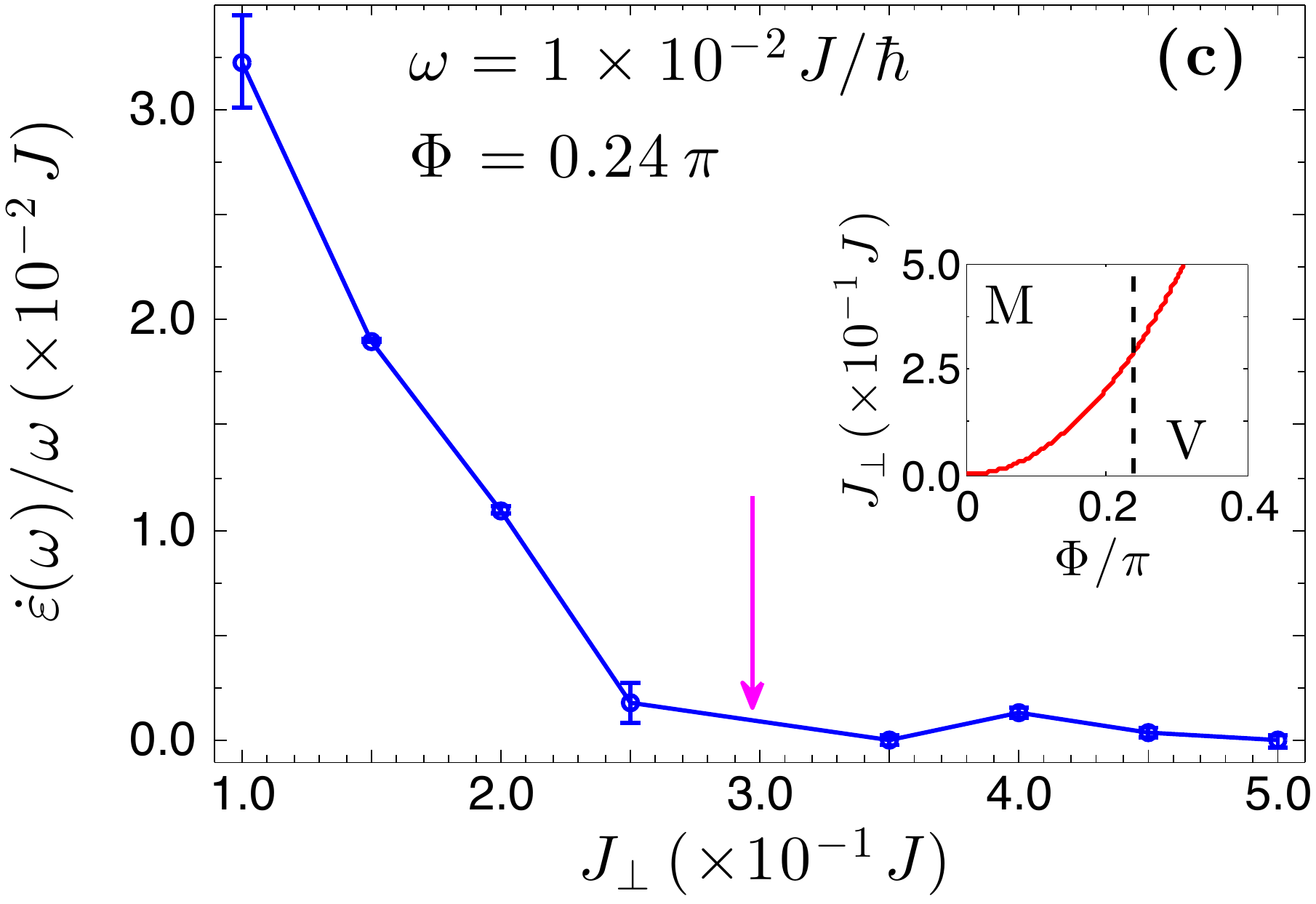}
\includegraphics[width=7.7cm]{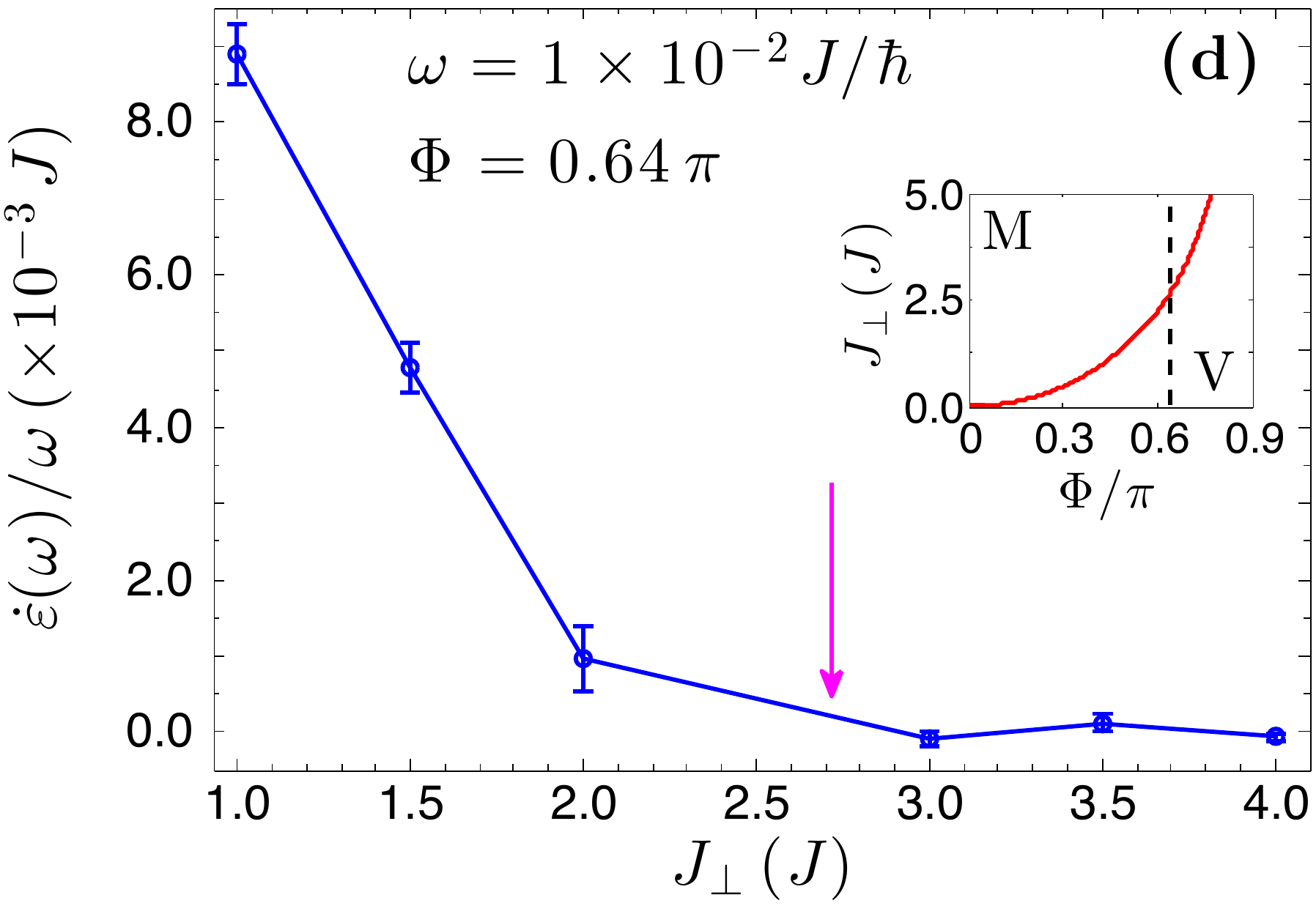}
\includegraphics[width=7.7cm]{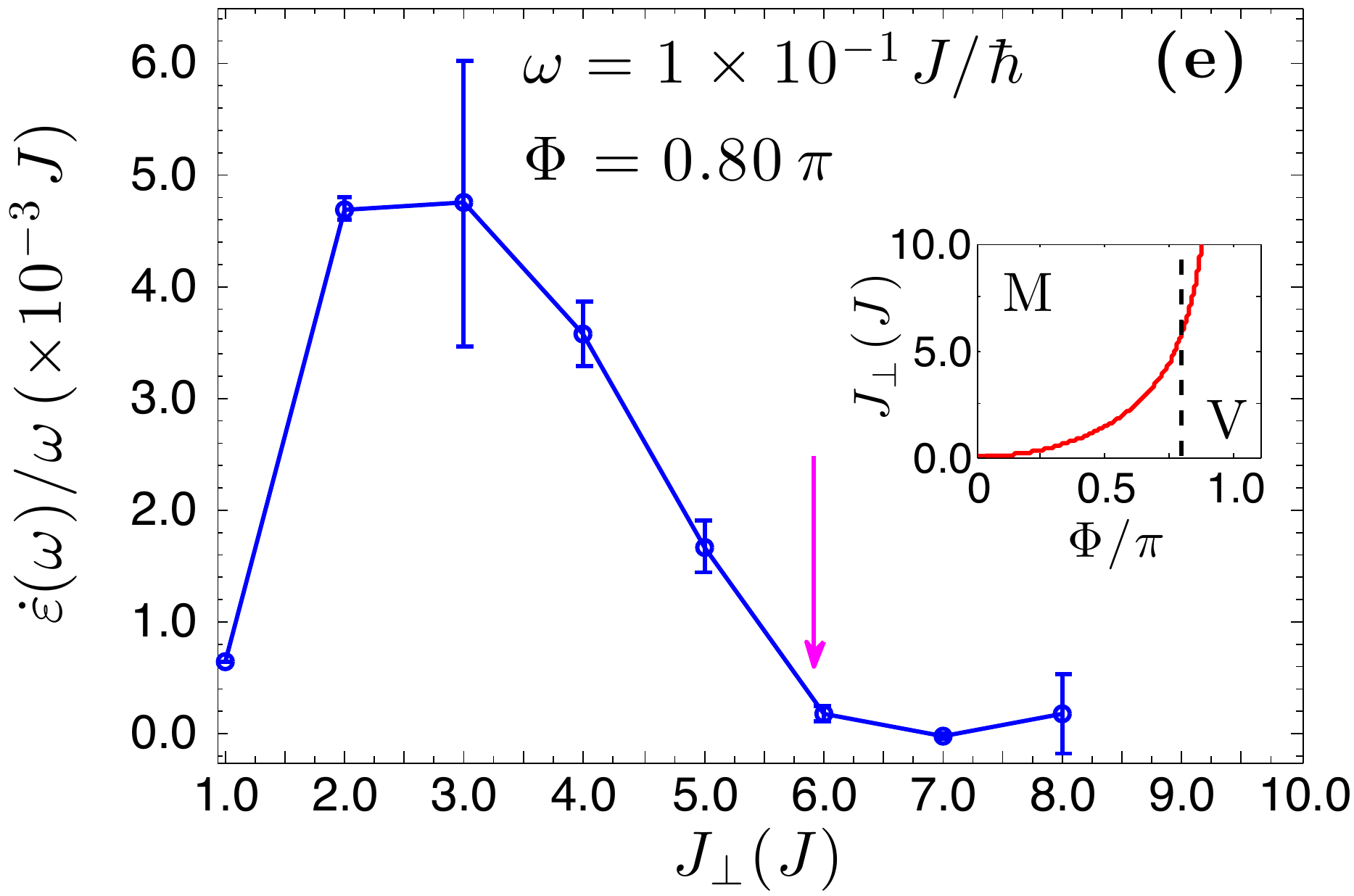}
\includegraphics[width=7.55cm]{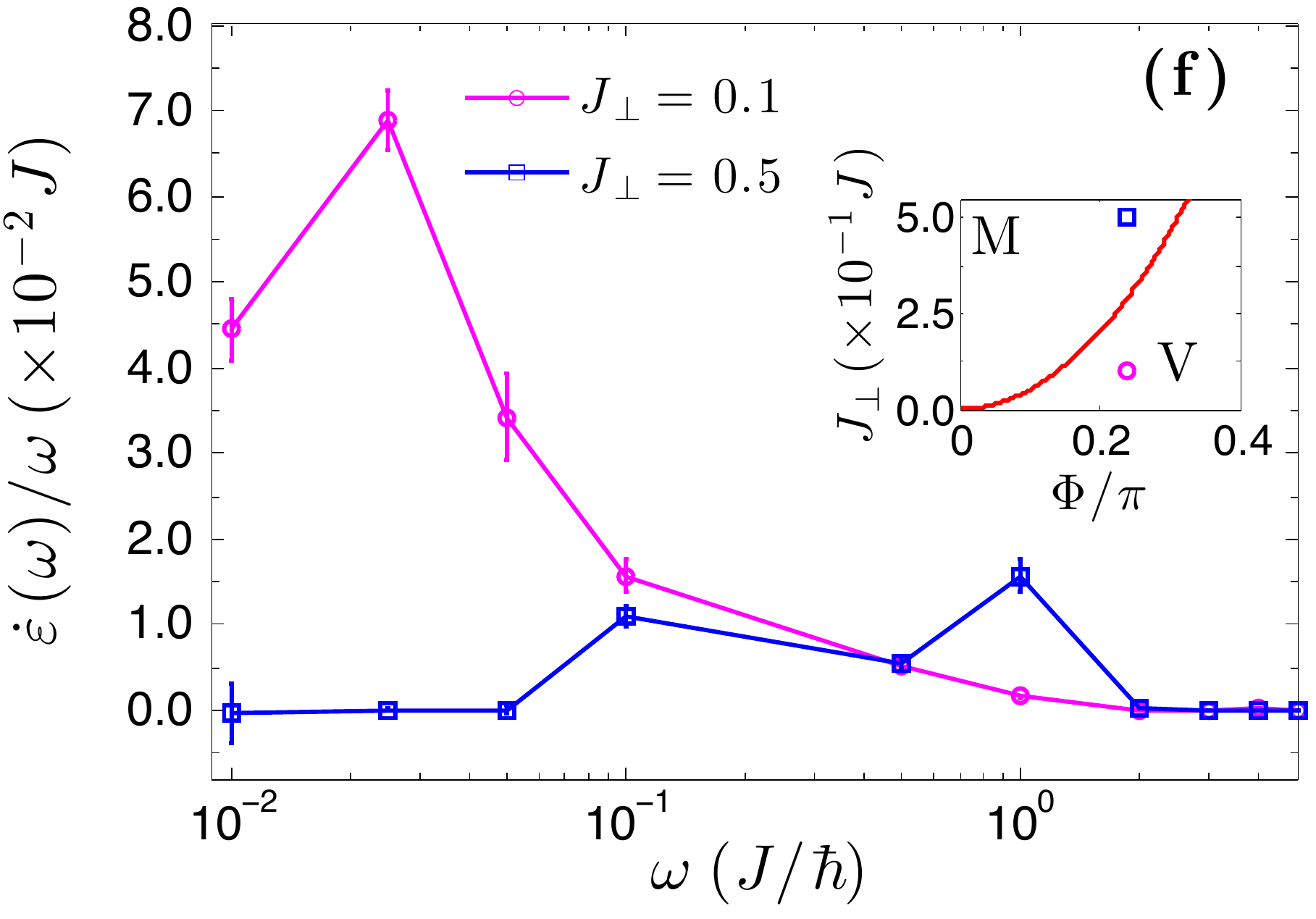}
\caption{Energy absorption for dilute bosons ($n =1/12 \ll 1)$. This observable monitors the response to a modulation of the spin imbalance with amplitude $\delta_1=0.4\,J$ and frequency $\omega$ as shown in the legends. \textbf{(a)} Relative energy variation, $\Delta E(t)=E(t)-E(0)$ in the V phase ($J_\perp/J=2.0$, magenta full line) and in the M phase ($J_\perp/J=7.0$, blue full line) for $\Phi/\pi=0.80$. The energy absorption rate (EAR) {per unit frequency} and its errors bars are extracted from a linear fit to $\Delta E(t)$ (black dashed line, see text for details). \textbf{(b)}-\textbf{(e)} EAR {per unit frequency} $\dot\varepsilon(\omega)/\omega$ as a function of $J_\perp$ for \textbf{(b)}:\,$\Phi/\pi=0.16$, \textbf{(c)}: \,$\Phi/\pi=0.24$, \textbf{(d)}:\, $\Phi/\pi=0.64$ and \textbf{(e)} $\Phi/\pi=0.80$. The insets in \textbf{(b)}-\textbf{(f)} show the non-interacting phase diagram ($U=0$) for reference. The V-M phase transition for non-interacting bosons are indicated by the magenta arrows in the main plots. \textbf{(f)} EAR {per unit frequency} as a function of $\omega$ for $\Phi/\pi=0.24$ in the V (magenta data) and in the M phase (blue data). For all plots reported here, the simulations were done for {hard-core} bosons using $L=24$ and $N=2$.}
\label{fig:eardilutecase}
\end{figure}

We first analyze a very dilute gas ($n\ll 1$) where interaction effects are weak and the physics is expected to be close to the free case{, for which} the critical line is analytically known~\cite{1367-2630-16-7-073005}. {We focus on the limit of hard-core bosons ($U\rightarrow\infty$). Differently from the equilibrium results presented in Section~\ref{sec:modelandobservables}, the increased numerical complexity of simulating the time evolution forces us to restrict ourselves to smaller values of the system size, namely $L=24$.}

The results are shown in Figure~\ref{fig:eardilutecase}, for $L=24$, $n=1/12$ and different values of $\Phi$. The amplitude of the density modulation is $\delta_1=0.4\,J$, and we use different values of $\omega$, depending on the value of $\Phi$. {During the time evolution, the time-dependent Hamiltonian in Equation~\eqref{eq:hamiltonianintheexperimentalgaugewiththeperiodicperturbation} is taken constant within each Trotter step. Therefore, the time step $dt$ has to be chosen small enough to ensure the reliability of this approximation for all the values of $\omega$ that we consider. We have verified that we can choose the time step $dt=10^{-2}\,\hbar/J$ in the Trotter expansion (see also~\ref{sec:appendix2} for a deeper discussion)}. In Figure~\ref{fig:eardilutecase}\textbf{(a)}, we show the relative energy variation, $\Delta E(t)=E(t)-E(0)$ for the two typical cases. In the M phase ($J_\perp/J=7.0$), there is no net energy absorption for sufficiently small $\omega$, whereas the system absorbs energy for all $\omega$ in the V phase ($J_\perp/J=2.0$). The EAR is extracted from the slope of $\Delta E(t)$ represented by the black dashed line. To remove the fast oscillations of $\Delta E(t)$ and extract the long-times linear trend, we perform  $M$ linear fits to $\Delta E(t)$ using different ranges of $t$. Accordingly, we obtain a set of values for the EAR {per unit frequency}, $\{\dot\varepsilon_q(\omega)/\omega\}_{q=1}^{M}$, from which we compute the mean value $\dot\varepsilon(\omega)/\omega=M^{-1}\sum_{q=1}^{M}\dot\varepsilon_q(\omega)/\omega$, and the standard deviation, $\sigma_{\dot\varepsilon}=\sqrt{M^{-1}\sum_{q=1}^{M}{[\dot\varepsilon_q(\omega)/\omega-\dot\varepsilon(\omega)/\omega]}^2}$. We take the latter as a measure of the uncertainty on the determined slope.

In Figure\,\ref{fig:eardilutecase}\textbf{(b)}-\textbf{(e)}, we show the EAR {per unit frequency} as a function of $J_\perp$ for four different values of $\Phi$. In the insets, we show the non-interacting phase diagram ($U=0$), where the red solid line corresponds to the critical line $J_{\perp,c}(U=0)$, and the black dashed line indicates the line at which we are cutting the phase diagram. The {behaviour} of the EAR {per unit frequency} is in agreement with the opening of the spin gap at the expected value of $J_{\perp,c}$. The system absorbs energy in the V phase ($J_\perp\lesssim J_{\perp,c}$), whereas the energy absorption ceases as the V-M phase transition takes place ($J_\perp\gtrsim J_{\perp,c}$). 

When $J_\perp=0$, the relation $[\hat N_s,\hat H_0]=0$ holds, implying that $[\hat N_s(t),\hat N_s]=0$ and $\chi_{N_s-N_s}(\omega)=0$ from Equation~\eqref{eq:responsefunction}. This is consistent with the curve plotted in Figure\,\ref{fig:eardilutecase}\textbf{(e)}, which tends to $0$ for low values of $J_\perp$. A similar {behaviour} is also expected for the values of $\Phi$ used in Figure\,\ref{fig:eardilutecase}\textbf{(b)}-\textbf{(d)}, but the considered value of $J_\perp$ was not small enough to highlight it.

As we previously stated, in the presence of a spin gap, the system is expected to absorb energy only if $\hbar\omega>\Delta E_s$. In Figure\,\ref{fig:eardilutecase}\textbf{(f)}, we show the EAR {per unit frequency} as function of $\omega$ for $\Phi/\pi=0.24$, both in the V phase ($J_\perp/J=0.1$) and in the M phase ($J_\perp/J=0.5$). For low modulation frequencies, we observe that the system can absorb energy in the V phase for values of the modulation frequency down to $\omega\sim10^{-2}\,J/\hbar$. In contrast, in the M phase, energy absorption starts from a finite frequency threshold, the value of which can be considered as a qualitative estimate of the spin gap. For high frequencies $\omega$, one observes a drop of the response in both phases, as expected from the general {behaviour} of the susceptibility $\chi(\omega)$ \cite{giamarchi2003quantum}.

\subsection{Results for strongly interacting gases}
\label{sec:resultforthemanybodycase}

We now move to the discussion of the strongly correlated case. {To approach this regime}, we consider {hard-core} bosons ($U\rightarrow\infty$) {and higher density with respect to the case in Section~\ref{sec:resultsforthedilutegascase}}. Previously, in Section~\ref{sec:themodel}, we showed how the presence of interactions shifts the critical point for the V-M phase transition. We here demonstrate that this shift is also detected by the periodic modulation of the density imbalance. 
{For concreteness, w}e focus on ${n=1/4}$ and $\Phi=0.8\,\pi${, as for the data in Figure~\ref{fig:meissnerandvortexphases}\textbf{(a)}}.

\begin{figure}[t]
\centering
\includegraphics[width=7.7cm]{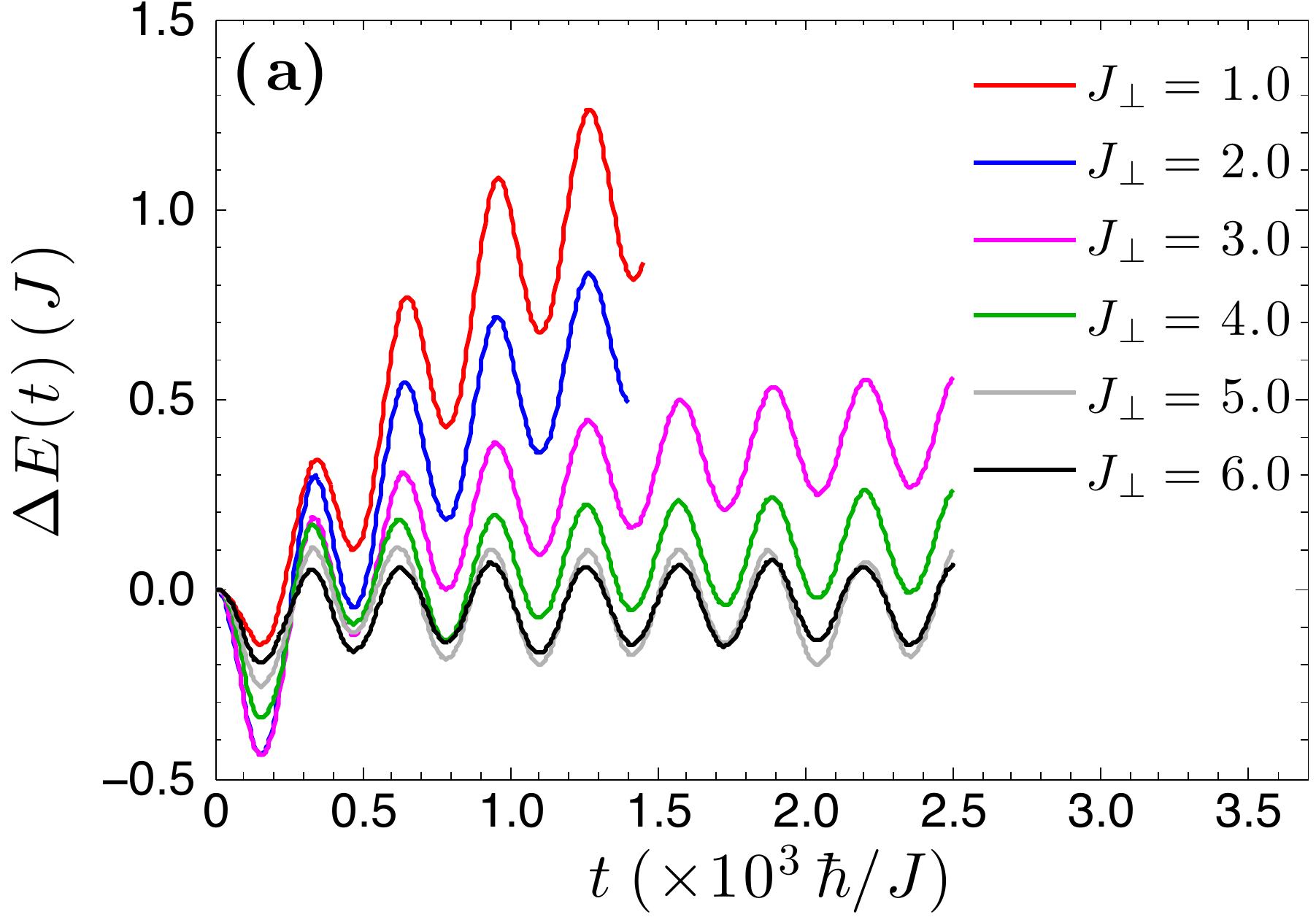}
\includegraphics[width=7.7cm]{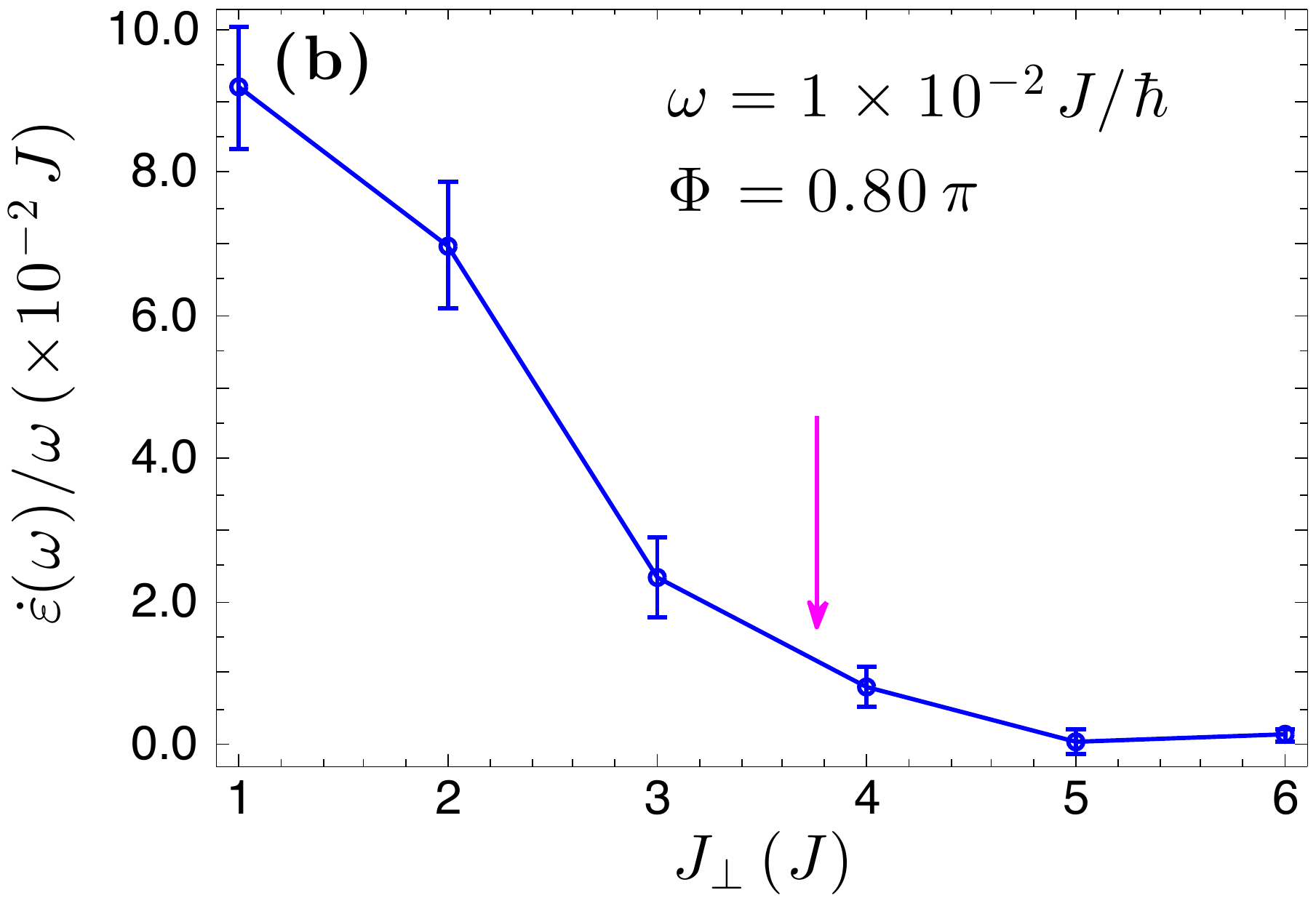}
\includegraphics[width=7.7cm]{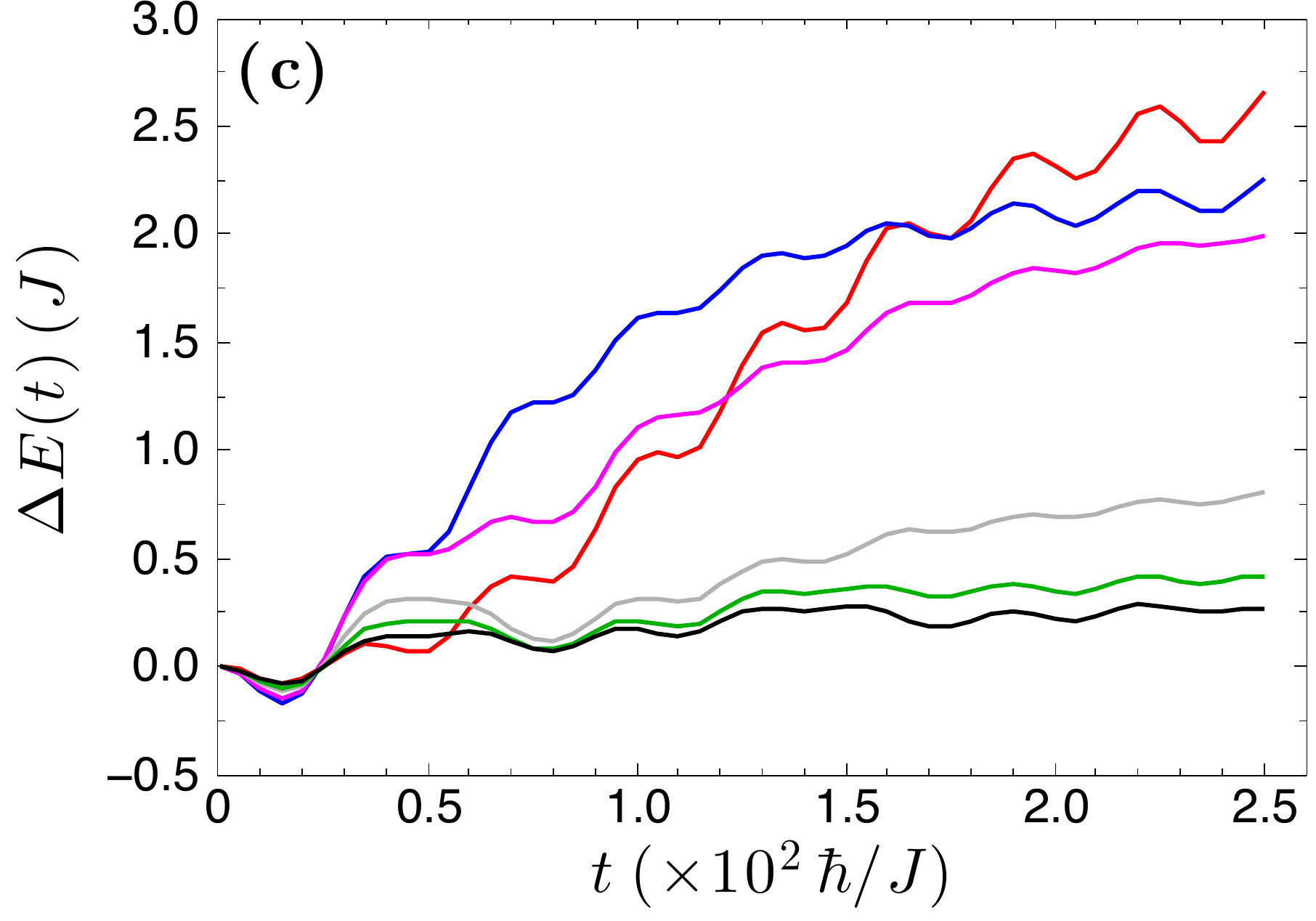}
\includegraphics[width=7.7cm]{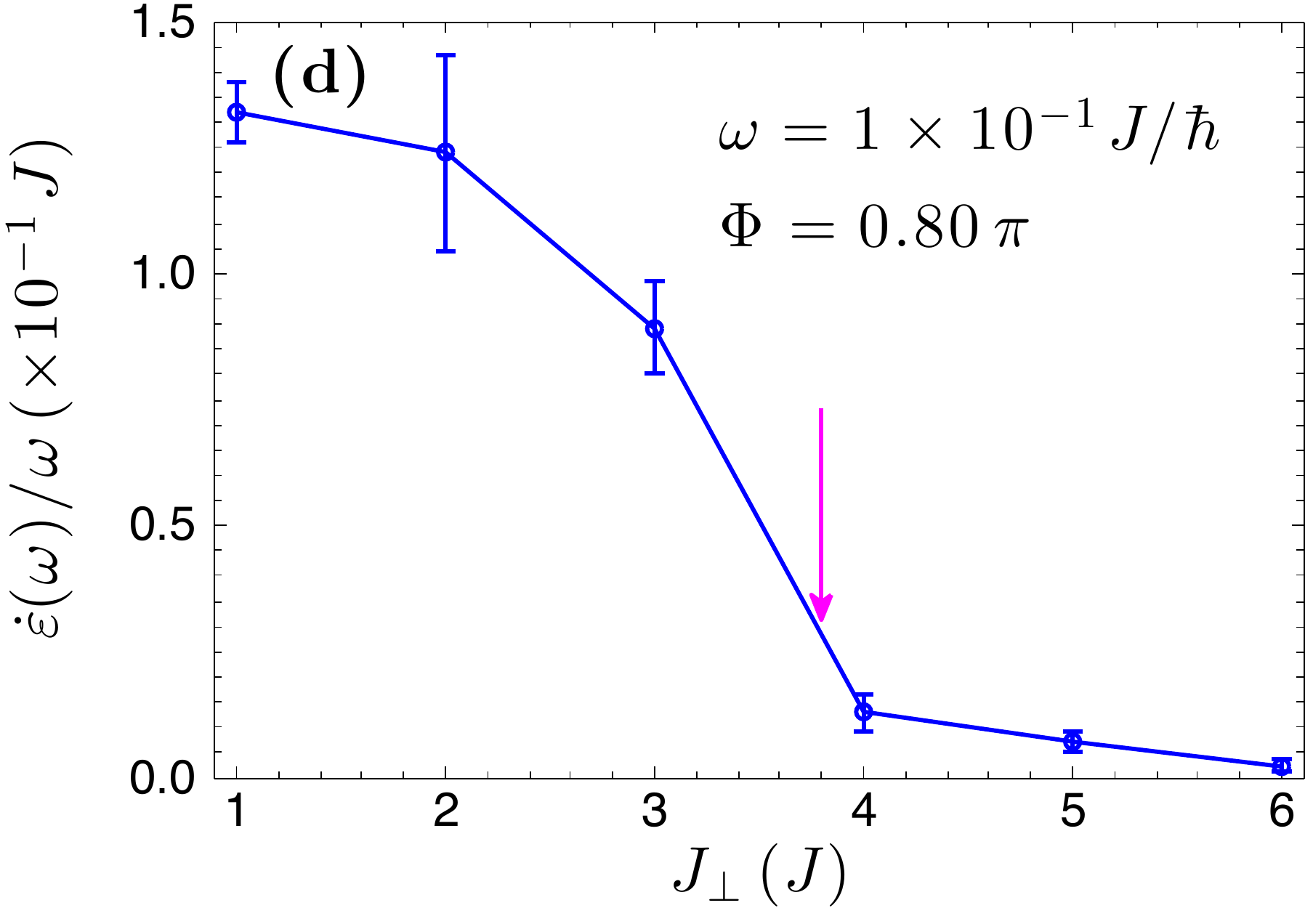}
\caption{Energy absorption for hard-core bosons with ${n=1/4}$. \textbf{(a)}: Energy absorption $\Delta E(t)$, and \textbf{(b)}: EAR {per unit frequency} for a modulation frequency $\omega=10^{-2}\,J/\hbar$. The magenta arrow indicates the estimated position of the M-V phase transition for hard-core bosons, $J_{\perp,c}/J\simeq3.8$ (see \ref{sec:appendix2}). \textbf{(c)} and \textbf{(d)} : Energy absorption and EAR {per unit frequency} for $\omega=10^{-1}\,J/\hbar$. In this case, we see a less sharp decreasing of the energy absorption for $J_\perp/J\gtrsim4.0$. For all plots, the spin imbalance is modulated with amplitude $\delta_1=0.4\,J$ and we use $L=24$ and $N=6$. The data at $J_\perp/J=1.0,2.0$ in panels \textbf{(a)}, \textbf{(b)} are obtained using a bond length $D_{{\rm max},t}=300$, whereas the other data are taken using $D_{{\rm max},t}=200$.  For panels \textbf{(c)}, \textbf{(d)}, we use {$D_{{\rm max},t}=350$}.}
\label{fig:earmanybodycase}
\end{figure}

The numerical results are shown in Figure~\ref{fig:earmanybodycase}. 
In Figure~\ref{fig:earmanybodycase}\textbf{(a)}, we plot $\Delta E(t)$ for different values of $J_\perp/J$ using  $\delta_1=0.4\,J$ and $\omega=10^{-2}\,J/\hbar$. In Figure~\ref{fig:earmanybodycase}\textbf{(b)}, we display the EAR {per unit frequency} as a function of $J_\perp/J$ for the same set of data. We use the same smoothing procedure as in the previous Section~\ref{sec:resultsforthedilutegascase}. The EAR {per unit frequency} vanishes for $J_\perp \geq 5.0 J $, and becomes nonzero when $J_\perp=4.0J$ and below. In Section~\ref{sec:themodel}, we estimated the critical value for the V-M transition, $J_{\perp,c}/J\simeq3.8$, slightly lower than the observed threshold for energy absorption. This quantitative discrepancy may be due both to finite-size effects, and to the fact that $\hbar\omega$ is possibly larger than the spin gap for $J_\perp/J = 4.0$.

In panels \textbf{(c)} and \textbf{(d)}, we show the same analysis for a larger value of $\omega$, namely $\omega=10^{-1}\,J/\hbar$. For $J_\perp/J\lesssim4.0$, the system absorbs energy {until saturation starts to take place}. Instead, for $J_\perp/J\gtrsim4.0$, energy absorption is suppressed. Differently from the data in panels \textbf{(a)} and \textbf{(b)}, we see a nonzero energy absorption also for $J_\perp/J=5.0,6.0$. We ascribe this fact to the larger value of $\hbar \omega$, possibly overcoming the value of the spin gap. The numerical complexity of the problem prevents us to use lower values of $\omega$, as the required simulation times $t$ are beyond our numerical possibilities. For a more critical discussion of the numerical data, see \ref{sec:appendix2}.

Concluding, our results are compatible with the opening of a spin gap around $J_\perp/J\simeq4.0$, which is in qualitative agreement with the phase diagram presented in Section~\ref{sec:themodel} for hard-core bosons and ${n=1/4}$. We thus conclude that the protocol we propose provides an experimentally accessible way to detect and measure the spin gap in the bosonic ladder all the way from the weakly to the strongly interacting regime.


\subsection{Discussion}
\begin{figure}[t]
\centering
\includegraphics[width=7.9cm]{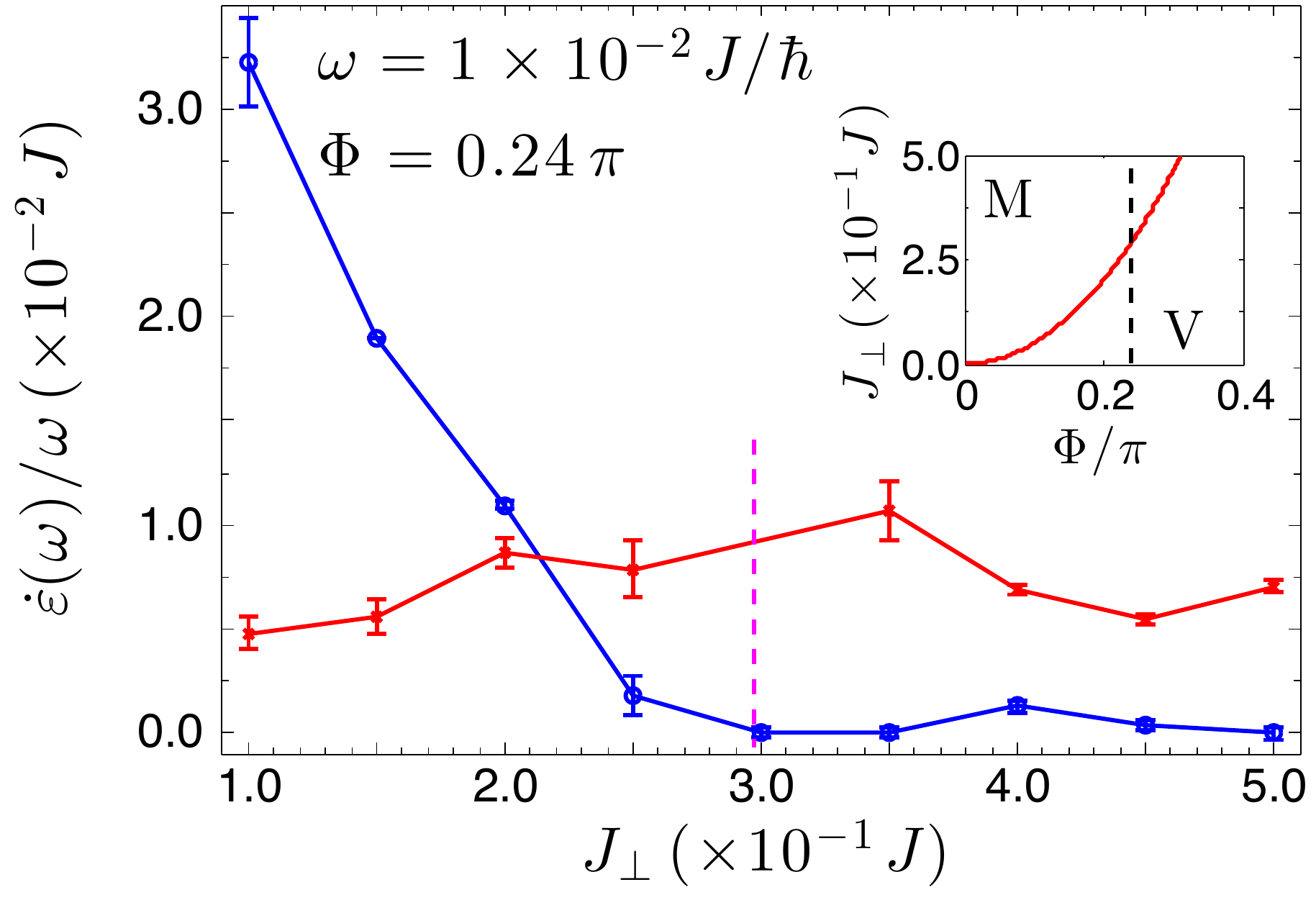}
\caption{{EAR per unit frequency} $\dot\varepsilon(\omega)/\omega$ as a function of $J_\perp$ for a dilute gas ($n=1/12$) at $\Phi=0.24\,\pi$. Other parameters are as in Figure~\ref{fig:eardilutecase}{\textbf{(c)}}. The blue or red points are obtained by modulating the spin imbalance $\hat N_s$ or the spin current $\hat{\mathcal{J}}_s$, respectively. The magenta vertical dashed line marks the M-V transition point $J_{\perp,c}$ for non-interacting bosons. The spin current is insensitive to the opening of the spin gap, and energy absorption takes place with roughly equal strength in both M and V phases.}
\label{fig:spingap}
\end{figure}

We conclude this Section with a discussion of the choice of the perturbation used to probe the system. Modulating the spin imbalance $\hat{N}_s$ is a natural choice to probe the properties of the system in the spin sector from an experimental perspective (see Section\, \ref{sec:experimentalrealizability}). As pointed out in Section~\ref{sec:modelandobservables}, both the M and V phases are gapless, and thus the choice of the modulation is crucial to distinguish them, since a generic one will in principle be sensitive to the presence of the gapless excitations and thus lead to absorption in both cases.

To display a counter-example, we show an additional calculation where the perturbation leads to energy absorption irrespective of whether the system is in the M or V phase. Instead of the spin density [Equation~\eqref{eq:hamiltonianintheexperimentalgaugewiththeperiodicperturbation}], we perturb the system using the perturbation $\hat V=F(t)\,\hat{\mathcal{J}}_s$, where the longitudinal spin-current operator $\hat{\mathcal{J}}_s$ is defined as
\begin{eqnarray}
\hat{\mathcal{J}}_s\equiv\sum_j\hat{\mathcal{J}}_{s,j}=\sum_j\left(\hat{\mathcal{J}}_{j,+\frac{1}{2}}-\hat{\mathcal{J}}_{j,-\frac{1}{2}}\right) \,\, .
\label{eq:spincurrentoperator}
\end{eqnarray}
In Equation~\eqref{eq:spincurrentoperator}, $\hat{\mathcal J}_{j,m}$ is the current operator on the link between site $j$ and $j+1$, and on the leg $m$:
\begin{eqnarray}
\hat{\mathcal J}_{j,m}=-iJ\left(\hat b^\dag_{j,m}\hat b_{j+1,m}-{\rm H.c.}\right) \,\,.
\end{eqnarray}
The results of the simulation are shown in Figure~\ref{fig:spingap}. We use the same system parameters as in Figure~\ref{fig:eardilutecase}\textbf{(c)}. The blue points correspond to the data for $\dot\varepsilon(\omega)/\omega$, as a function of $J_\perp$, when the system is modulated by using $\hat N_s$, whereas the red point correspond to $\dot\varepsilon(\omega)/\omega$ when $\hat{\mathcal J}_s$ is instead used. As we show in the figure, when we perturb the system using $\hat{\mathcal{J}}_s$, energy absorption takes place both in the V and in the M phase. Thus, the choice of using the spin current as a perturbation does not allow us to probe the spin gap, differently from the case when the spin density is used.


\section{Experimental realization using laser-induced tunneling}
\label{sec:experimentalrealizability}

As discussed in the Introduction, most experimental realizations of bosonic flux ladders with cold atoms do not strictly realize the situation described by the Hamiltonian in Equation~\eqref{eq:bosonictwolegladderhamiltonian} due to different interaction terms. In the approach of \cite{NatPhys10.588.14}, the interaction energy per atom is very weak due to the large number of atoms per site, and in \cite{science1514}, interactions are long-ranged in the synthetic (spin) dimension. The bosonic FL with strong, short-range interactions, but only for two particles, has been also investigated~\cite{nature22811}. Here, we discuss an alternative experimental realization that follows from the proposal of~\cite{njp.12.033007} for realizing the Harper-Hofstadter Hamiltonian in a square optical lattice. This scheme naturally realizes a bosonic FL with short-range (on-site) interactions and low filling around or below one atom per site.

\begin{figure}[t]
\centering
\includegraphics[width=0.8\textwidth]{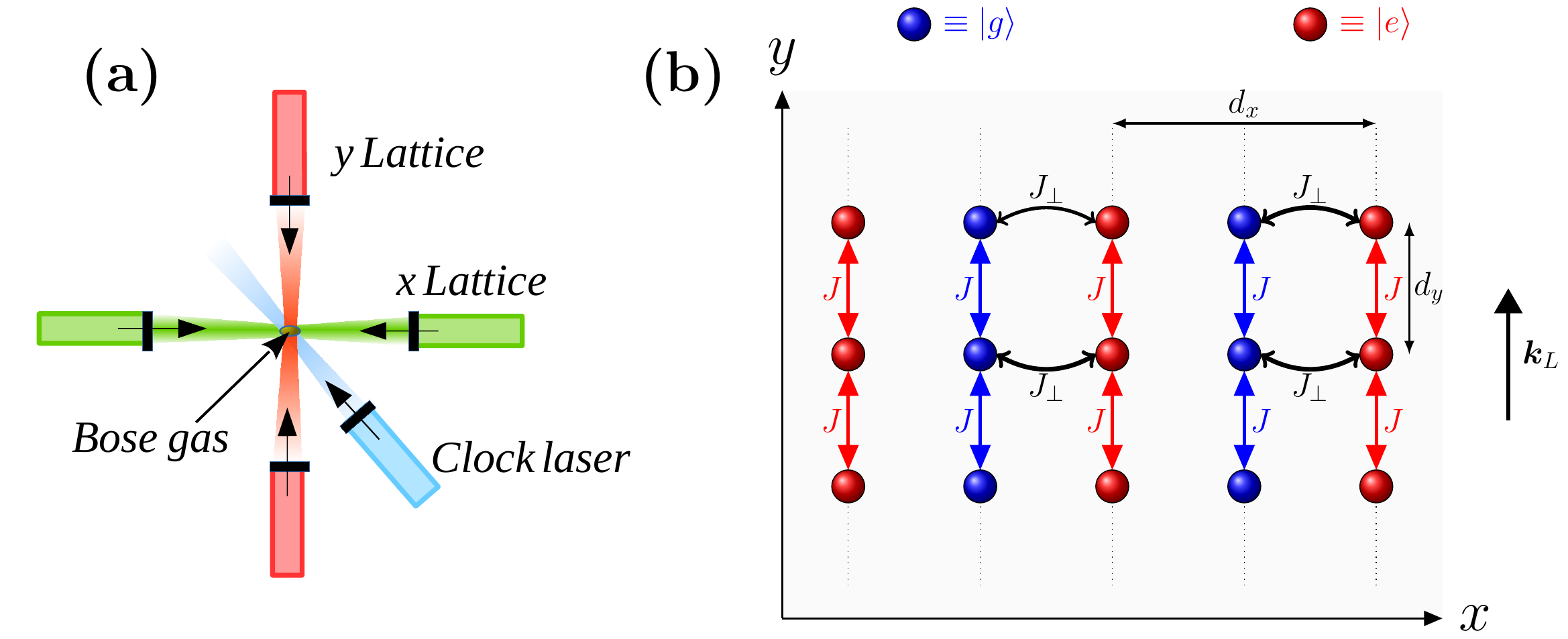}
\caption{A possible experimental realization following \cite{njp.12.033007}. \textbf{(a)}: Sketch of the laser arrangement. \textbf{(b)}: Spin-dependent lattice realizing a collection of disconnected two-leg ladders.}
\label{fig:exp_figure}
\end{figure}

We first review the scheme described in~\cite{njp.12.033007}. We consider an atomic species with two long-lived internal states connected by an ultra-narrow optical transition as used in optical atomic clocks~\cite{RevModPhys.87.637}. This can be realized, \textit{e.g.} using the singlet $^1$$S_0$$= g$ GS and a metastable $^3$$P_0$$= e$ state in group-II or Ytterbium atoms. The atoms are trapped in two dimensions by a strong confining potential along $z$, and in the $x-y$ plane by a state-dependent square optical lattice trapping atoms in different sublattices depending on their internal state (see Figure\,\ref{fig:exp_figure} and \cite{1367-2630-10-7-073015,njp.12.033007}). The $y$ lattice of period $d_y$ is chosen to trap atoms in both internal states identically. The $x$ potential is formed by the sum of a short lattice with spacing $d_x$, $V_{x,\mu}(x)=\epsilon_\mu V_{0,x} \cos^2(\pi x/d_x+\phi_{\rm SL})$, with $\mu=g,e$ and with $\epsilon_\mu=+1$ for $g$ and $-1$ for $e$, and of a long lattice with spacing $2 d_x$, $W_{\mu}(x)=W_\mu \cos^2(\pi x/2d_x+\phi_{W})$, with a well-controlled relative phase $\phi_{W}$ \cite{foelling2007a}. By suitably choosing the depths of the $x$ lattices, one can suppress standard tunneling along $x$ within each sublattice $g$ or $e$.

A laser of wavevector $k_L$ is then used to coherently couple states $g$ and $e$, thereby inducing hopping between the $g$ and $e$ sublattices. This laser-assisted tunneling process \cite{ruostekoski2002a,jaksch2003a} is described by a tight-binding Hamiltonian of the form~\eqref{eq:bosonictwolegladderhamiltonian} with 
\begin{eqnarray} \label{eq:phi_exp}
\Phi = \frac{\mathbf{k}_L \cdot \mathbf{d}_y}{2\pi},
\end{eqnarray}
with $\mathbf{d}_y=d_y \mathbf{e}_y$. For Ytterbium atoms, for instance, $d_y \simeq 380\,$nm and $2\pi/k_L\simeq 578\,$nm, leading to a maximum value of $\Phi_{\rm max}\simeq 0.66$ when the coupling laser propagates along $y$. The value of $\Phi$ can be tuned between $0$ and $\Phi_{\rm max}$ by changing the direction of propagation of the laser. A calculation of the band structure leads to laser-induced tunneling energies of $J_\perp/h\sim 100$\,Hz for $V_{0,x}=8\,E_{\rm r,x}$ and $W_{\mu} \ll V_{0,x}$, where $E_{\rm r,x}/h \simeq 3\,$kHz is the recoil energy associated with the period-$d_x$ lattice \cite{njp.12.033007}. Note that $J_\perp$ is proportional to the power of the coupling laser, and that the intra-leg tunneling $J$ is tuneable independently by changing the depth of the $y$ lattice.

The simultaneous presence of the superlattice and laser coupling enlarges the unit cell to $2 d_x$, with in general four non-equivalent sites per unit cell (two associated with $g$ and two with $e$). This corresponds to four different types of $g-e$ ``links'' and to four different transition frequencies, which are non-degenerate for a generic $\phi_W$. By a suitable choice of $\phi_{W}$, two of these links can be made degenerate \cite{njp.12.033007}. Connecting all neighboring lattice sites with resonant laser-assisted tunneling then requires three different transition frequencies $\omega_1, \omega_1\pm W/\hbar$ (where $W$ is related to the amplitudes $W_e,W_g$). Choosing $W$ large enough compared to the laser-induced tunneling energies $J_\perp$ ensures that a given laser frequency only enables tunneling for the links where it is resonant (typically one can choose $W/h\sim 8$\,kHz and $W/J_\perp \sim 80$). This setup leads to a two-dimensional Hofstadter optical lattice with a uniform flux $\Phi$ through each unit cell. This fully connected Hofstadter lattice can be reduced in a straightforward manner to an array of two-leg ladders by removing every other frequency $\omega_1 \pm W/\hbar$ (see Figure\,\ref{fig:exp_figure}\textbf{(b)}). Similarly, three-leg ladders could be realized by removing only one frequency, for instance $\omega_1+W/\hbar$. 

Focusing now on the two-leg ladder geometry, each leg of the ladders is associated with a different internal state $g \equiv +1/2$ or $e\equiv -1/2$. In this situation, time-of-flight and state-dependent imaging (see, \textit{e.g.},~\cite{goldman2012a}) gives access to the leg-resolved MDF. Furthermore, a non-zero detuning $\delta_1=\omega_1-\omega_{eg}$ of the coupling laser from the atomic resonance $\omega_{eg}$ generates a term $\propto \hat N_s$, as desired for the spectroscopy protocol presented in Section~\ref{sec:spectroscopy}. Frequency modulation of $\omega_1$ is straightforward to implement using acousto-or electro-optical modulators, and energy absorption can be detected by monitoring the changes of the MDF.


\section{Conclusions}\label{sec:conclusions}

In this article, we have investigated the properties of bosonic flux ladders from the dilute to the strongly correlated regime. For particle densities $n <1$, the phase transition from a Meissner to a vortex phase is qualitatively unchanged, but quantitatively strongly affected by interactions. With the help of numerical simulations, we have shown that this phase transition can be observed by recording the momentum distribution, and that its precise location is well identified by the ``imbalance ratio'' characterizing the multi- or single-peak character of the momentum distribution.

{Moreover, we have discussed a spectroscopic method that employs a periodic modulation of the spin imbalance between the two legs as a probe of the excitation spectrum. Gapped spin-like excitations in the Meissner phase prevent energy absorption below a certain frequency threshold, that we identified with the spin gap; in contrast, energy absorption occurs at all frequencies in the vortex phase. As such, monitoring the energy absorbed versus the modulation frequency allows one to measure not only the location of the phase transition, but also the value of the spin gap. 

The characterization of the low-energy properties of a quantum many-body system is as important as the characterization of the state itself.
Since we have shown that the protocols discussed in this article are within the reach of state-of-the art experiments, 
we believe that our work will motivate further interest in the study of the low-energy properties of complex quantum phases by indicating an effective procedure to be applied in the non-trivial cases where gapped and gapless excitations of different nature coexist. 


\section*{Acknowledgements}
We {thank} J. Beugnon, M. Bosch Aguilera, R. Bouganne, S. De Palo, and R. Fazio for fruitful discussions. We are also grateful to D. Rossini for providing the MPS code {and for support}. L. M. was supported by LabEX ENS-ICFP: ANR-10-LABX-0010/ANR-10-IDEX-0001-02 PSL*. This work was granted access to the HPC resources of MesoPSL financed by the Region Ile de France and the project Equip@Meso (reference ANR-10-EQPX-29-01) of the programme Investissements d'Avenir supervised by the Agence Nationale pour la Recherche. We also acknowledge the CINECA award under the ISCRA initiative, for the availability of high performance computing resources and support.


\appendix
\section{Numerical analysis of the phase diagram}
\label{sec:appendix1}
In this appendix, we discuss our results on the phase diagram obtained in Figure~\ref{fig:meissnerandvortexphases}. For sufficiently large $L$, the phase transition from the V to the M phase can be numerically detected by computing the central charge, which is extracted from the \emph{entanglement entropy} (EE). The EE is defined by $S(\ell)=-{\rm Tr}\left[\hat\rho_\ell\log\left(\hat\rho_\ell\right)\right]$, $\hat\rho_\ell$ being the reduced density matrix of a bipartition of the chain of length $\ell$. In the case of OBC, the leading {behaviour} of the EE computed on the GS is predicted to be~\cite{1742-5468-2004-06-P06002}
\begin{eqnarray}
S(\ell)=s_1+\frac{c}{6}\log\left[\left(\frac{2L}{\pi}\right)\sin\left(\frac{\pi\ell}{L}\right)\right] \,\, ,
\label{eq:entanglemententropycalabresecardy}
\end{eqnarray}
where $s_1$ is a non-universal value and $c$ is the central charge, which gives the number of gapless modes in the system. Thus, for $n<1$, one predicts $c=2$ in the V phase, and $c=1$ in the M phase, where the spin sector is gapped~\cite{PhysRevB.91.140406}.

\begin{figure}[t]
\centering
\includegraphics[width=7.7cm]{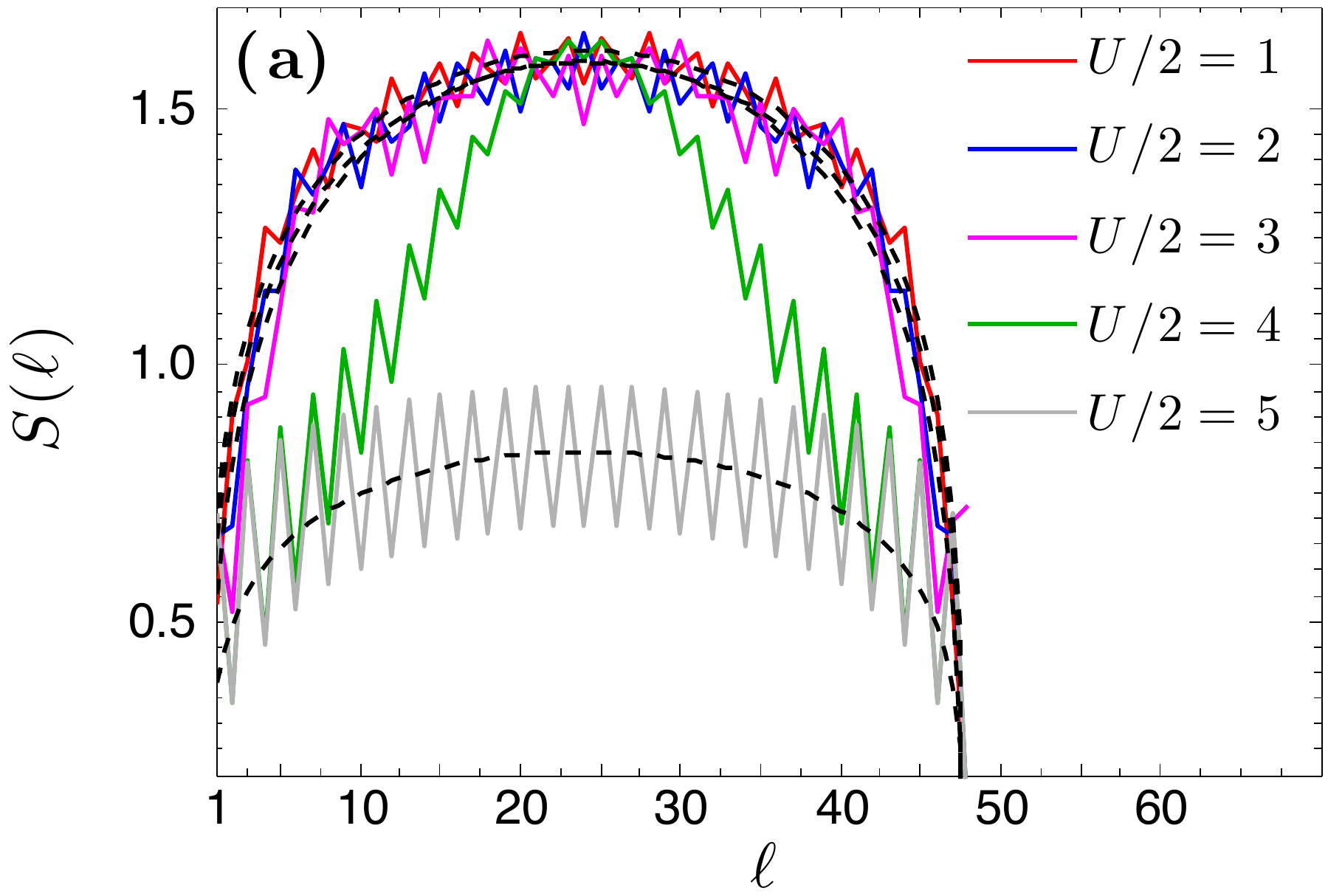}
\includegraphics[width=7.7cm]{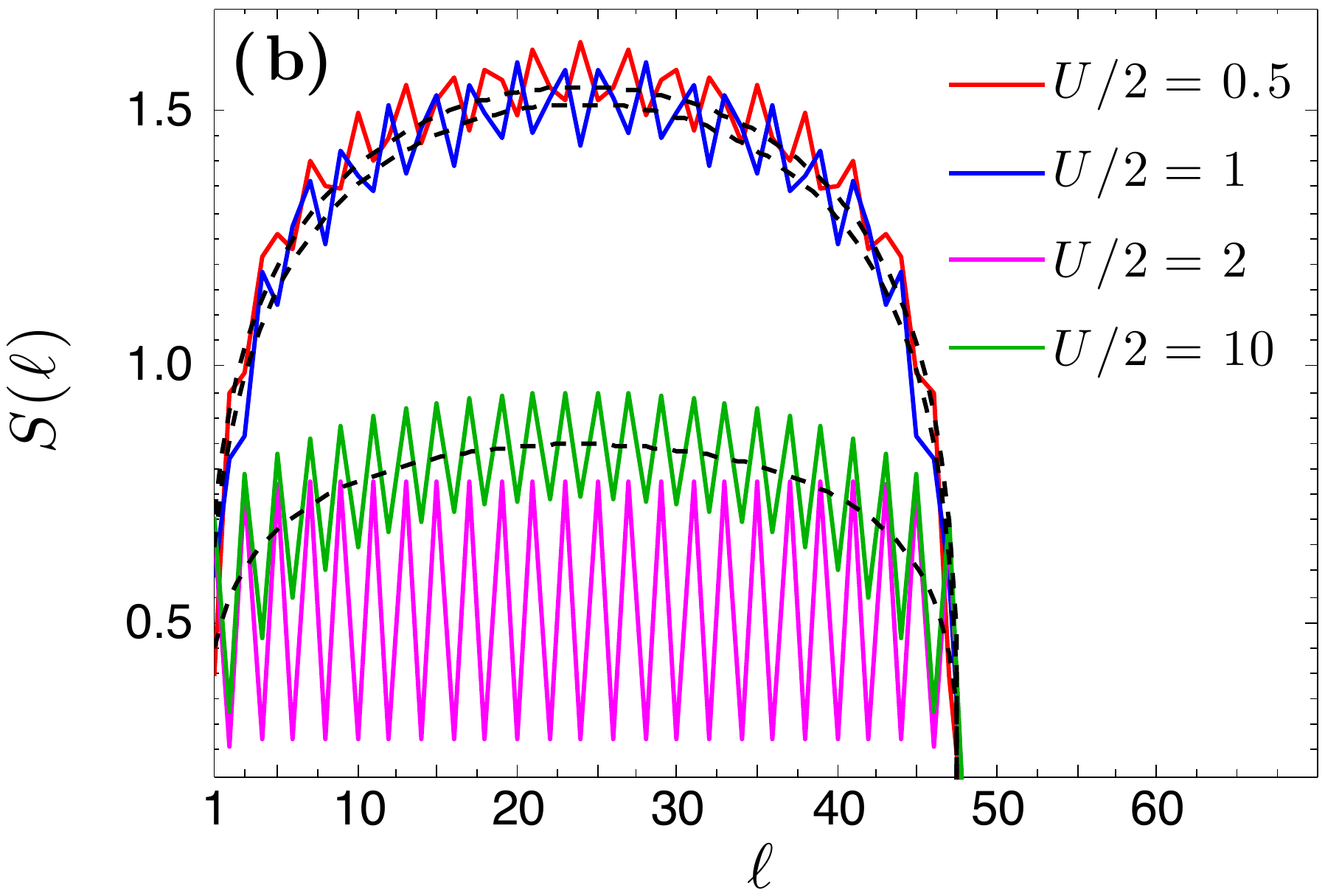}
\includegraphics[width=7.7cm]{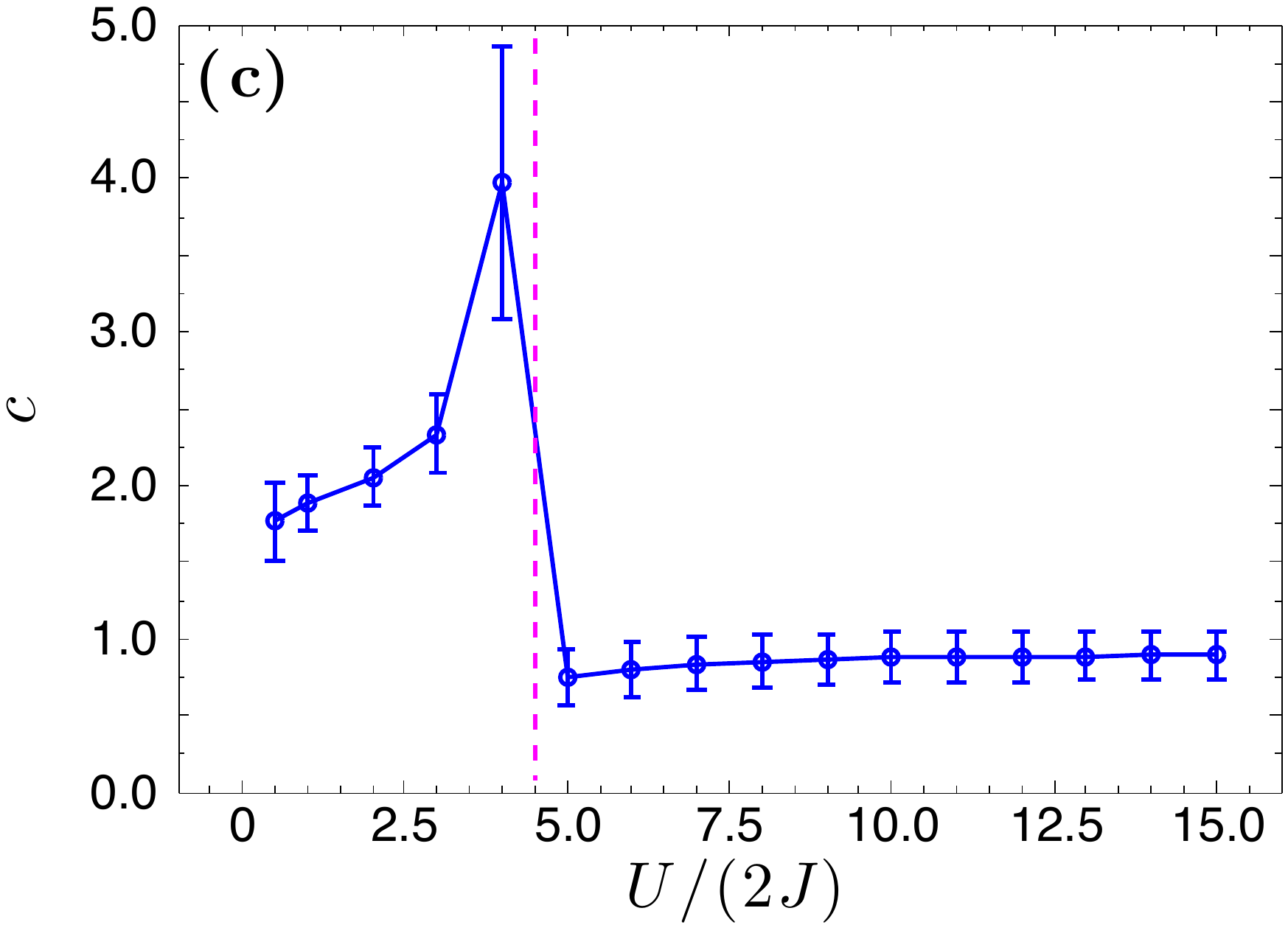}
\includegraphics[width=7.7cm]{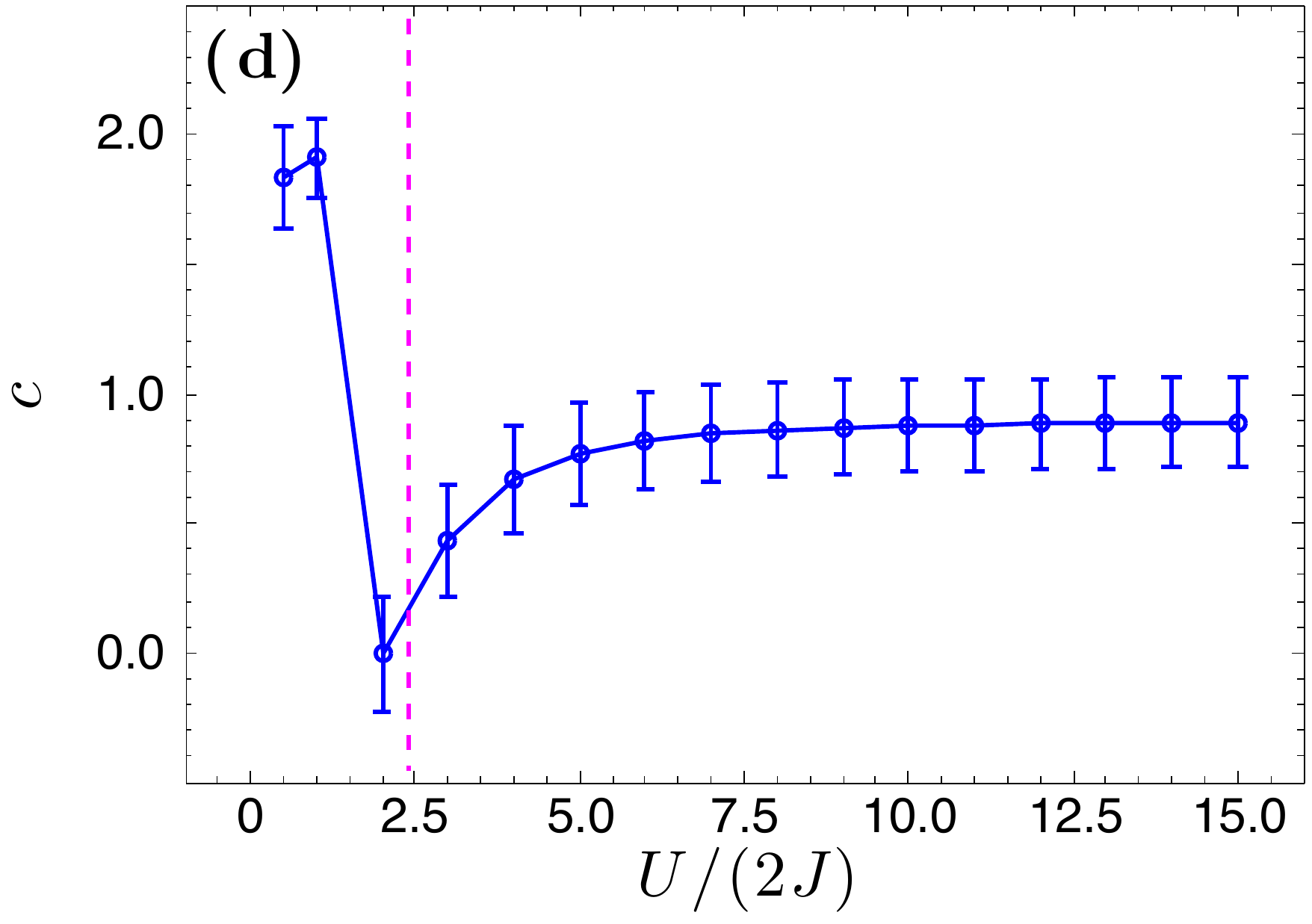}
\caption{Analysis of the EE for the phase diagram in Figure~\ref{fig:meissnerandvortexphases} for ${n=1/2}$. We show \textbf{(a)} the EE for $J_\perp/J=1.75$ and \textbf{(b)} for $J_\perp/J=2.00$, for different values of $U/J$ across the V-M phase transition. The phase transition is detected from the sudden change of the EE. Such a change is well reflected by the central charge, which is shown as a function of $U/(2J)$ for \textbf{(c)} $J_\perp/J=1.75$ and \textbf{(d)} $J_\perp/J=2.00$. The uncertainties are estimated as explained in the text. Sufficiently far away from the transition point, the values of $c$ that we fit are in agreement with the expected ones. The transition points estimated from the jump of the central charge are in agreement with what we found in Figure~\ref{fig:meissnerandvortexphases}\textbf{(b)} by looking at the IR (magenta dotted line).}
\label{fig:eeanalysis}
\end{figure}

The analysis of the EE and of the central charge for the data of the phase diagram in Figure~\ref{fig:meissnerandvortexphases}\textbf{(b)}, with ${n=1/2}$, is reported in Figure~\ref{fig:eeanalysis}. In panels \textbf{(a)} and \textbf{(b)}, we show the EE for different values of $U/(2J)$ as in the legends, across the V-M phase transition (see Figure~\ref{fig:meissnerandvortexphases}\textbf{(b)}, magenta line). We perform a fit with Equation~\eqref{eq:entanglemententropycalabresecardy} (black dashed lines in Figure\,\ref{fig:meissnerandvortexphases}\textbf{(b)}) to extract the central charge. Close to the V-M phase transition, Equation~\eqref{eq:entanglemententropycalabresecardy} fails to describe the {behaviour} of the EE, but sufficiently far away from the transition point the fit agrees well with the numerical data. {Such behaviour of the EE has been observed also in other models~\cite{PhysRevB.94.115112,PhysRevB.87.115132,PhysRevB.92.035154,PhysRevX.7.021033,arXiv:1707.05715}, and ascribed to the fact that, in the vicinity of the phase transition, the low-energy excitations become massive because of the presence of a gapped low-energy spectrum, and the leading order of $S(\ell)$ is not described by Equation~\eqref{eq:entanglemententropycalabresecardy} any more.}

We show the central charge as a function of $U/(2J)$, for the same set of data, in panels \textbf{(c)} and \textbf{(d)}. We ascribe the fact that we do not fit exactly $c=1$ or $c=2$ to finite-size effects. Because of the oscillatory {behaviour} of the EE and of the choice of OBC, to fit the EE and compute the values of $c$ for each value of $U/(2J)$, we repeat the fit $N_c$ times, introducing a cutoff $L_c$ which we vary from $L_c=1$ to $L_c=N_c$. For each repetition of the fit, we fit including only points in the range $\ell\in[L_c:L-L_c]$. We accordingly obtain a set of values for the central charge, $\{c_{L_c}\}_{L_c=1}^{N_c}$, from which we estimate the mean value as $\bar c={N_c}^{-1}\sum_{L_c=1}^{N_c}c_{L_c}$, and the uncertainty by means of the standard deviation $\sigma_c=\sqrt{{N_c}^{-1}\sum_{L_c=1}^{N_c}{(c_{L_c}-\bar c)}^2}$. As we see from the figure, the phase transition from the V phase to the M phase is identified by the jump of the central charge. Furthermore, sufficiently far away from the V-M transition point, the fitted values of $c$ are in agreement with the expected values predicted by bosonization.  We conclude by noting that the transition points estimated from the IR in Figure~\ref{fig:meissnerandvortexphases} are in agreement with the one estimated from the numerically determined central charge, the latter being known to signal the M-V phase transition~\cite{PhysRevB.91.140406}.

\begin{figure}[t]
\centering
\includegraphics[width=7.7cm]{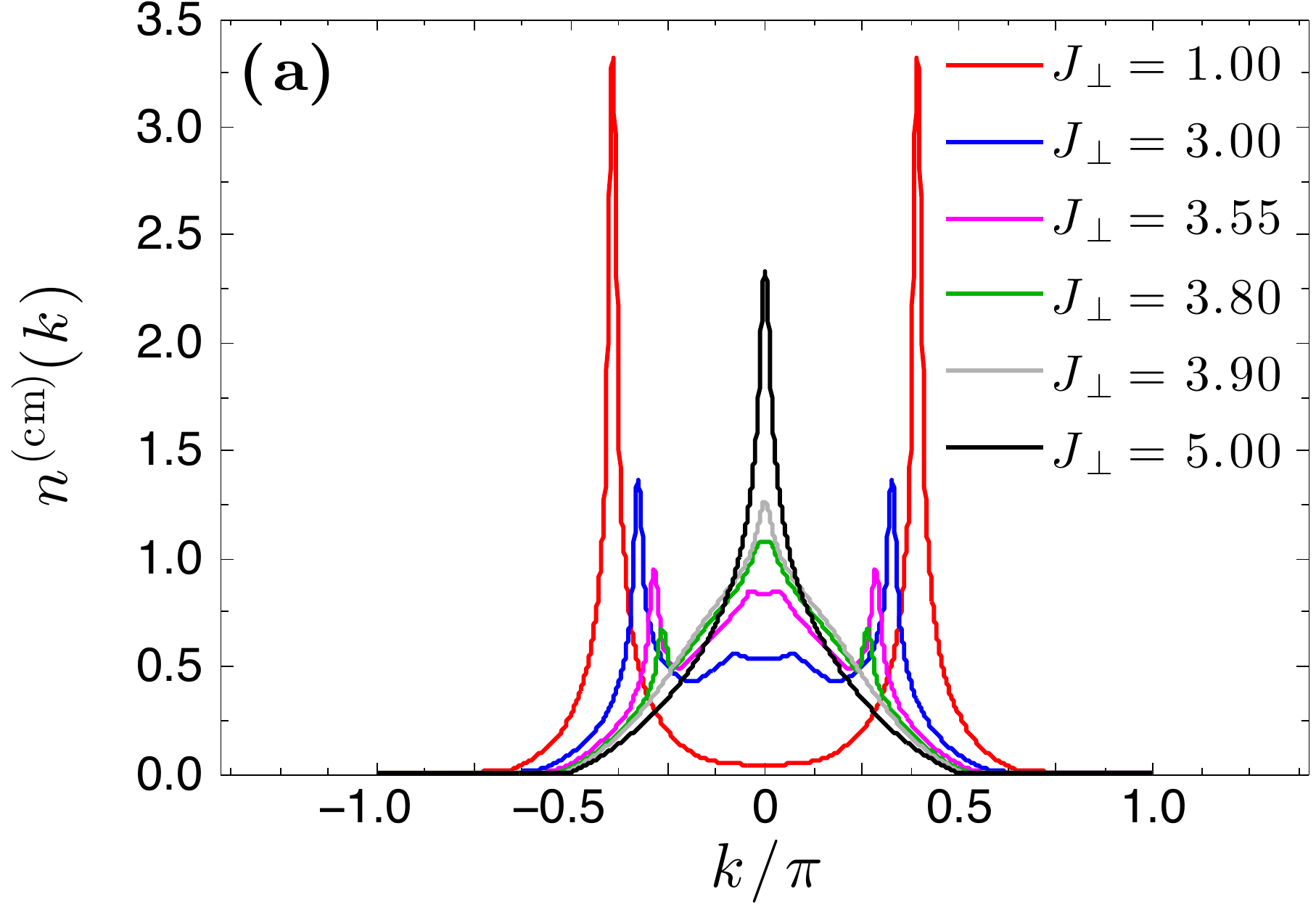}
\includegraphics[width=7.5cm]{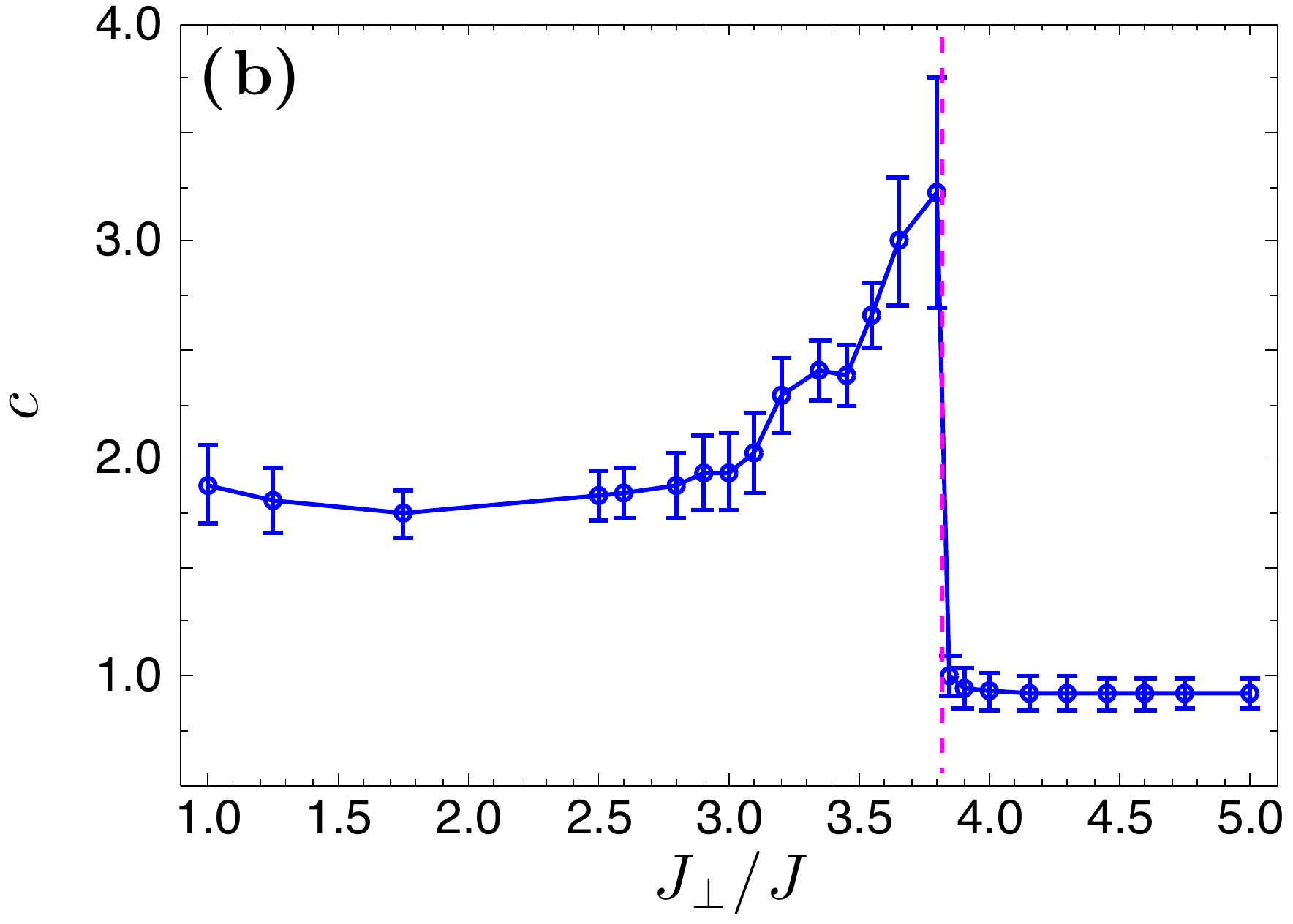}
\caption{Analysis of the V-M phase transition on the $U/J\rightarrow\infty$ line of the phase diagram in Figure~\ref{fig:meissnerandvortexphases}\textbf{(a)}. We simulate HCBs using ${n=1/4}$ and $L=96$. We show \textbf{(a)} the total MDF in the condensed-matter gauge for different values of $J_\perp/J$ across the V-M phase transition, and \textbf{(b)} the central charge as a function of $J_\perp/J$, computed as explained in the text. We see that the total MDF becomes single-peaked around $J_\perp/J\simeq3.8$ (see also Figure~\ref{fig:meissnerandvortexphases}\textbf{(a)}). Accordingly, the central charge drops down to $c=1$ around the same value of $J_\perp/J$, where the magenta dotted line indicates the value of $J_{\perp,c}/J$ found by looking at the IR. As in Figure~\ref{fig:eeanalysis}, the critical value agrees with the one estimated via the IR. Also, the {behaviour} of the EE fails to be described by Equation~\eqref{eq:entanglemententropycalabresecardy} (not shown) close to the transition point, and thus the values of $c$ {that} we fit deviate from the expected ones. Instead, sufficiently far away from the phase transition, the values of $c$ {that} we fit are in agreement with the expected values ($c=2$ in the V phase and $c=1$ in the M phase).}
\label{fig:hcbanalysis}
\end{figure}

The numerical simulations at finite $U$ are performed by truncating the local Hilbert space on each site $j$ and leg $m$, which we denote by $\mathcal{H}_{j,m}$. We define by $|r\rangle_{j,m}$ the local Fock space such that $\mathcal{H}_{j,m}={\rm span}\{|r\rangle_{j,m}\}_{r=0}^{d_{\rm loc}-1}$, where $d_{\rm loc}={\rm dim}(\mathcal{H}_{j,m})$. Let $\hat{\mathbb{P}}^{(j,m)}_r=|r\rangle_{j,m}\langle r|$ be the local projector over the state $|r\rangle_{j,m}$. The local density operator is then $\hat n_{j,m}=\sum_{r=0}^{d_{\rm loc}-1}r\,\hat{\mathbb{P}}^{(j,m)}_r$. The suitable choice for $d_{\rm loc}$ depends on the values of $U/(2J)$; we can keep up to $d_{\rm loc}$ states for $\mathcal{H}_{j,m}$ if the probability of finding $d_{\rm loc}-1$ particles on the site $j$ of the leg $m$ is small with respect to the local density, i.e. $\langle\hat{\mathbb{P}}^{(j,m)}_{d_{\rm loc}-1}\rangle\ll\langle\hat n_{j,m}\rangle$, for all $j$ and $m$, where the expectation value is computed on the GS of the system. Thus, we choose to verify that $L^{-1}\sum_j\langle\hat{\mathbb{P}}^{(j,m)}_{d_{\rm loc}-1}\rangle\ll L^{-1}\sum_j\langle\hat n_{j,m}\rangle$, for all $m$. In our numerical simulations, we see that this condition is fulfilled using $d_{\rm loc}=3$, for all the values of $U/(2J)$ that we use, since we verify that $L^{-1}\sum_j\langle\hat{\mathbb{P}}^{(j,m)}_{d_{\rm loc}-1}\rangle \lesssim \epsilon L^{-1}\sum_j\langle\hat n_{j,m}\rangle$, for all $m$, with $\epsilon=3\times10^{-2}$ a small numerical factor.

We now discuss the numerical estimation of the critical point for hard-core bosons at ${n=1/4}$ and $\Phi/\pi=0.8$. We have chosen $U\rightarrow \infty$ to simulate longer chains ($L=96$) and reduce finite-size effects while having a sufficiently low numerical complexity. We compute the total MDF in the condensed-matter gauge and the EE, from which we extract the central charge. The result is shown in Figure~\ref{fig:hcbanalysis}. Deep in the V phase, the MDF displays two symmetric peaks with respect to $k=0$. As the V-M phase transition is approached, additional peaks around $k=0$ start to appear, and one peak eventually dominates when one enters the M phase. The phase transition is also signaled by the jump of the central charge, which drops from $c=2$ in the V phase to $c=1$ in the M phase. The EE and central charge display the same {behaviour} as in the previous case, and are analyzed in the same way. We finally estimate $J_{\perp,c}(\infty,n)\simeq3.8\,J$ from the {behaviour} of the central charge, which agrees with the value we estimate by measuring the IR (Figure~\ref{fig:meissnerandvortexphases}\textbf{(a)}).


\section{Details on the {time-dependent} numerical calculations}
\label{sec:appendix2}
In this appendix, we discuss the effect of a finite value of the bond link $D_{\rm max,t}$ {and of the time step $dt$} in the numerical calculation using the TEBD algorithm. In order to ensure the reliability of our data for long times, the value of $D_{{\rm max},t}$ must be large enough to take into account the increasing amount of entanglement in the system, which is particularly important for the deep V phase. We first focus on the data in panels \textbf{(a)} and \textbf{(b)}, which are taken using $\omega=10^{-2}\,J/\hbar$, with $D_{{\rm max},t}=300$ (for $J_\perp/J=1.0,2.0)$ and $D_{{\rm max},t}=200$ (for $J_\perp/J\geq3.0)$. In our simulations, we see that the bond link $D_t$ starts to saturate to $D_{{\rm max},t}$ at the sites around $L/2$ after a time which is smaller than the total simulation time.

\begin{figure}
\centering
\includegraphics[width=7.7cm]{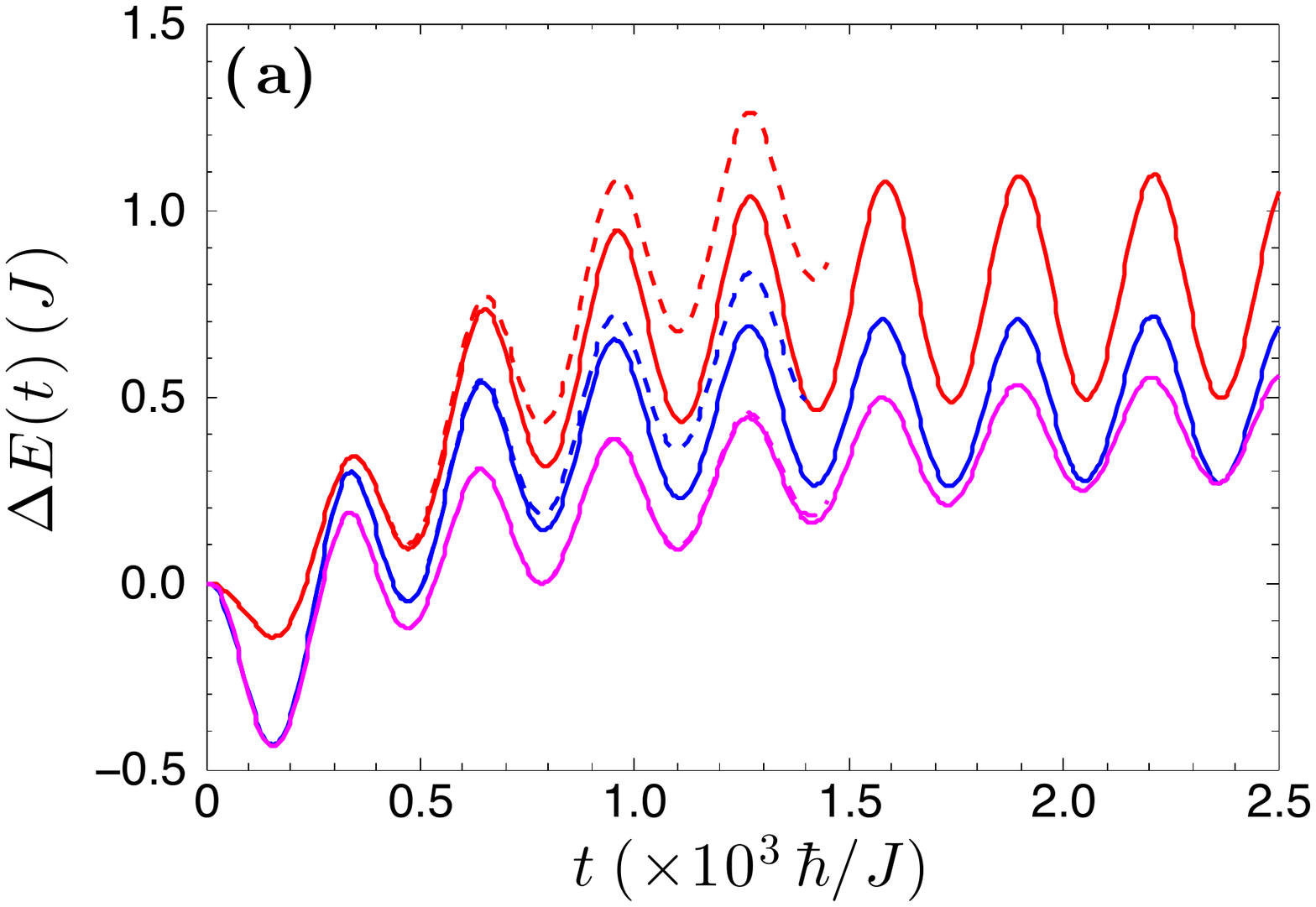}
\includegraphics[width=7.7cm]{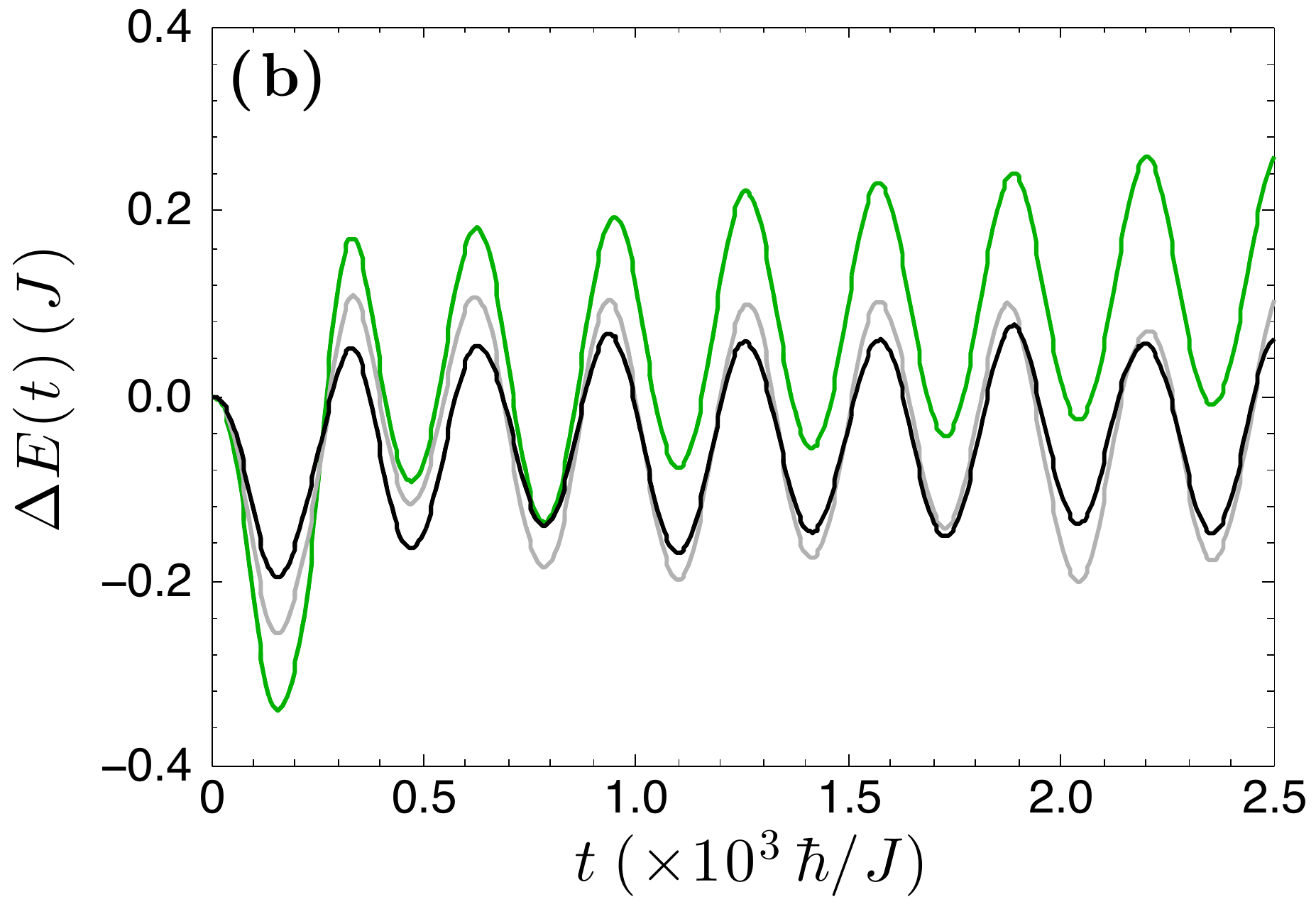}
\caption{Numerical data of $\Delta E(t)$ for the data in Figure~\ref{fig:earmanybodycase}\textbf{(a)}. We show $\Delta E(t)$ for \textbf{(a)} $J_\perp/J=1.0$ (red lines), $J_\perp/J=2.0$ (blue lines) and $J_\perp/J=3.0$ (magenta lines), and \textbf{(b)} $J_\perp/J=4.0$ (green lines), $J_\perp/J=5.0$ (grey lines) and $J_\perp/J=6.0$ (black lines). Solid lines are taken using $D_{{\rm max},t}=200$, whereas dashed lines are taken with $D_{{\rm max},t}=300$.}
\label{fig:bondlinkanalysis1}
\end{figure}

In order to see how this fact affects our data of $\Delta E(t)$, we compare the results for $\Delta E(t)$ by using $D_{{\rm max},t}=200$ and $D_{{\rm max},t}=300$. The result is shown in Figure~\ref{fig:bondlinkanalysis1}. In particular, we separately show $\Delta E(t)$ in the V phase (Figure~\ref{fig:bondlinkanalysis1}\textbf{(a)}) and in the M phase (Figure~\ref{fig:bondlinkanalysis1}\textbf{(b)}). The data at $D_{{\rm max},t}=200$ and $D_{{\rm max},t}=300$ are shown using solid and dashed lines respectively. As evident from the figure, the curves with $D_{{\rm max},t}=200$ become significantly different in the deep V phase ($J_\perp/J=1.0,2.0$) from the curves computed using $D_{{\rm max},t}=300$ for times which are between $t=500\,\hbar/J$ and $t=750\,\hbar/J$, i.e., after $D_t$ has saturated to $D_{{\rm max},t}=200$ almost on all sites of the chain. Indeed, in the V phase, where we have $c=2$, we see that the saturation of the bond link to $D_{{\rm max},t}=200$ starts after $t\simeq160\,\hbar/J$, for $J_\perp/J=1.0,2.0$, and after $t=370\,\hbar/J$ for $J_\perp/J=3.0$. Instead, in the M phase, where we have $c=1$, the bond link increases in time with a smaller rate with respect to the data in the V phase: for the data at $J_\perp/J=4.0$, we start to see saturation of the bond link to $D_{{\rm max},t}=200$ after $t\simeq1500\,\hbar/J$, whereas $D_t$ never saturates for $J_\perp/J>4.0$. Thus, from this analysis, we see that we need to use at least $D_{{\rm max},t}=300$ for the data at $J_\perp/J=1.0,2.0$, whereas we can use $D_{{\rm max},t}=200$ for the others.

\begin{figure}
\centering
\includegraphics[width=7.7cm]{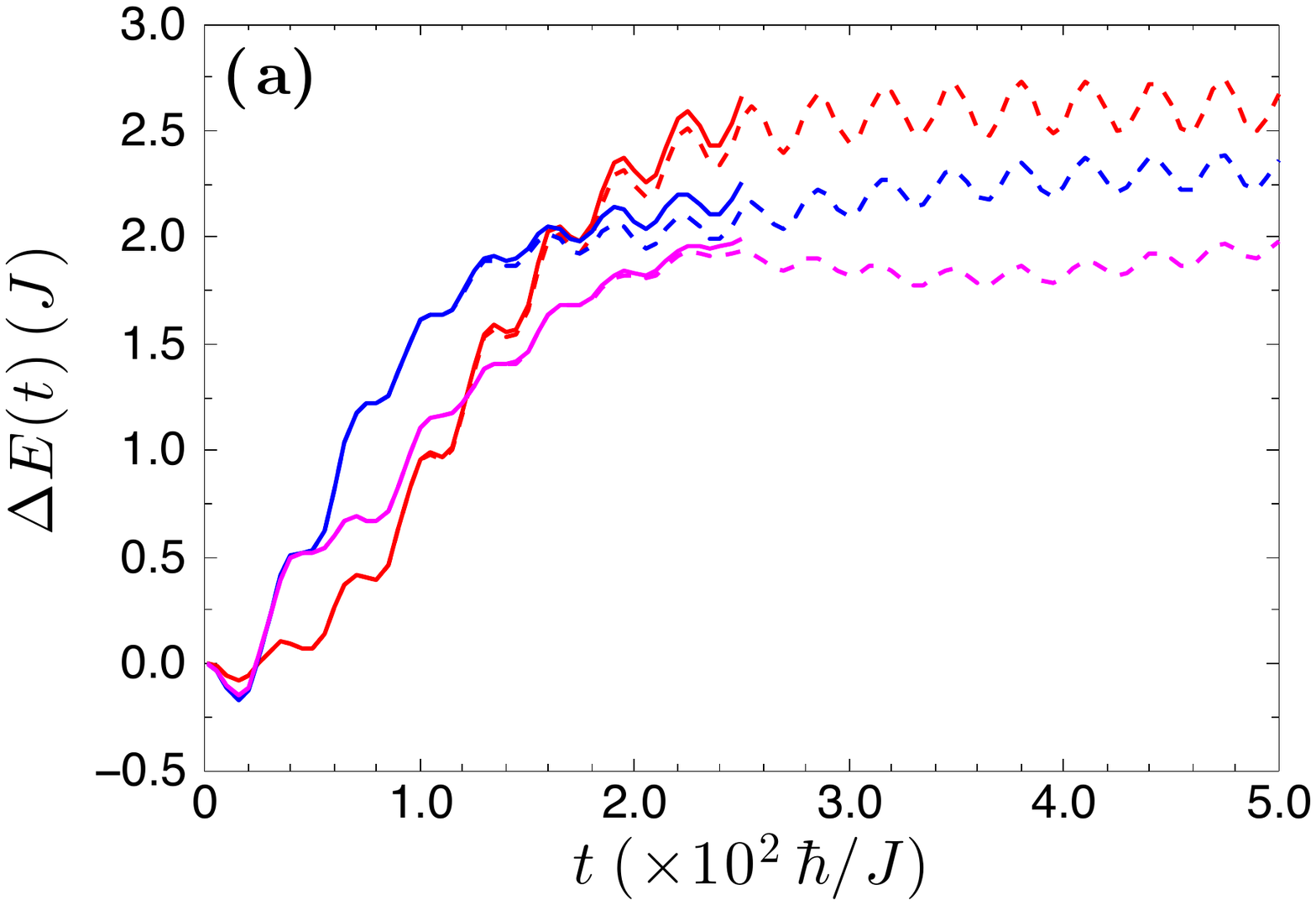}
\includegraphics[width=7.7cm]{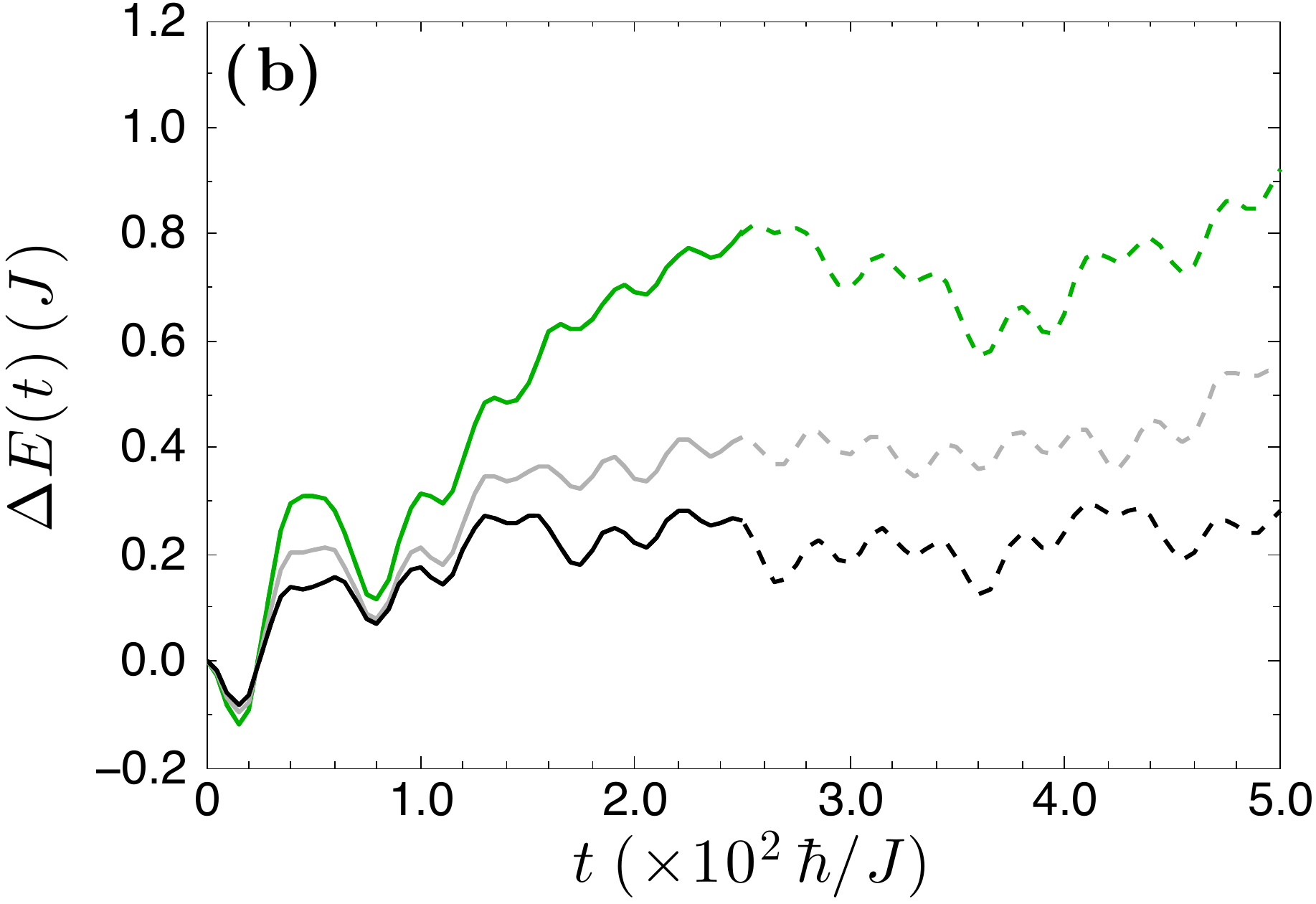}
\caption{Numerical data of $\Delta E(t)$ for the data in Figure~\ref{fig:earmanybodycase}\textbf{(c)}. We show $\Delta E(t)$ for \textbf{(a)} $J_\perp/J=1.0$ (red lines), $J_\perp/J=2.0$ (blue lines) and $J_\perp/J=3.0$ (magenta lines), and \textbf{(b)} $J_\perp/J=4.0$ (green lines), $J_\perp/J=5.0$ (grey lines) and $J_\perp/J=6.0$ (black lines). Solid lines are taken using $D_{{\rm max},t}=350$, whereas dashed lines are taken with $D_{{\rm max},t}=300$.}
\label{fig:bondlinkanalysis2}
\end{figure}

We perform the same analysis for the data in Figure~\ref{fig:earmanybodycase}\textbf{(c)}, which are taken at $\omega=10^{-1}\,J/\hbar$. In this case, we can simulate up to shorter times with respect to the case in Figure~\ref{fig:earmanybodycase}\textbf{(a)}. This allows us to use larger values of the bond link, which we choose {$D_{{\rm max},t}=350$}. In order to see the effect of the finite value of $D_{{\rm max},t}$, we then compare these data of $\Delta E(t)$ with the data computed using {$D_{{\rm max},t}=300$}. The result is shown in Figure~\ref{fig:bondlinkanalysis2}. In this case, we observe that $D_t$ starts to saturate to $D_{{\rm max},t}=350$ already at $t=25\,\hbar/J$ in the V phase. The fact the $D_t$ grows in time with a larger rate with respect to the case in Figure~\ref{fig:bondlinkanalysis1} is due to the larger value of $\omega$ {that} we use.

As in Figure~\ref{fig:bondlinkanalysis1}, the data in the M phase (Figure~\ref{fig:bondlinkanalysis2}\textbf{(b)}) are less sensitive to the bond link difference with respect to the data in the V phase because of the smaller amount of entanglement. As we found for the data in Figure~\ref{fig:bondlinkanalysis1}, we here see that the different values of $D_{{\rm max},t}$ during the TEBD algorithm do not drastically affect the qualitative {behaviour} of $\Delta E(t)$, and thus of the EAR, {at least for the times considered for the fits}.

{As we pointed out in Section~\ref{sec:resultsforthedilutegascase}, also the time step $dt$ has to be properly chosen in order to ensure the correct convergence of the TEBD algorithm. In our algorithm, during the time evolution and for each Trotter step, the Hamiltonian is taken constant within the time interval $dt$. Since the Hamiltonian depends explicitly on time through the function $F(t)$ [see Equation~\eqref{eq:hamiltonianintheexperimentalgaugewiththeperiodicperturbation}], it is important to check the validity of this approximation for the choice $dt=10^{-2}\,\hbar/J$, specifically in the large-$\omega$ limit considered in the data in Figure~\ref{fig:eardilutecase}\textbf{(f)}. To do so, we compared the results of the simulations in Figure~\ref{fig:eardilutecase}\textbf{(f)} with the results of a simulation with the same parameters but using $dt=10^{-3}\,\hbar/J$. We found that, $|\dot\varepsilon(\omega,dt=10^{-2}\,\hbar/J)/\omega-\dot\varepsilon(\omega,dt=10^{-3}\,\hbar/J)/\omega|/J\lesssim10^{-4}$ even for the largest values of $\omega$ that we consider (not shown), where $\dot\varepsilon(\omega,dt)/\omega$ indicates the data series of the EAR per unit frequency taken using the time step $dt$. Therefore, the choice of $dt=10^{-2}\,\hbar/J$ in the fourth-order Trotter expansion is sufficient to ensure the correct convergence of the TEBD algorithm}.


\section{{Additional data in the interacting regime}}
{We now extend the discussion carried out in Figure~\ref{fig:eardilutecase}\textbf{(e)} and Figure~\ref{fig:earmanybodycase}\textbf{(d)}. In the former case ($L=24$ and $N=2$), we showed that the energy absorption starts to be suppressed when $J_\perp/J\simeq6.0$, in agreement with the analytical result for free bosons $J_{\perp,c}(U=0)\simeq5.9\,J$ \cite{1367-2630-16-7-073005}, whereas in the latter case ($L=24$ and $N=6$), the suppression of the energy absorption was observed approximatively from $J_\perp/J\simeq4.0$. This result is in agreement with the critical point $J_{\perp,c}\simeq3.8\,J$ that we numerically estimated in Figure~\ref{fig:hcbanalysis}, in the case of a long chain ($L=96$) at filling ${n=1/4}$. In order to pinpoint the reliability of these numerical data and in order to ensure that the shift of the critical point, estimated from the spectroscopic method, that we observe from Figure~\ref{fig:eardilutecase}\textbf{(e)} to Figure~\ref{fig:earmanybodycase}\textbf{(d)} is not an artifact due to finite-size effect, we here show additional numerical data, simulating hard-core bosons with the same parameters as in Figure~\ref{fig:eardilutecase}\textbf{(e)} and Figure~\ref{fig:earmanybodycase}\textbf{(d)}, but using a different value of the density, $n=1/6$ (i.e., $N=4$ with $L=24$).}

\begin{figure}[t]
\centering
\includegraphics[width=7.85cm]{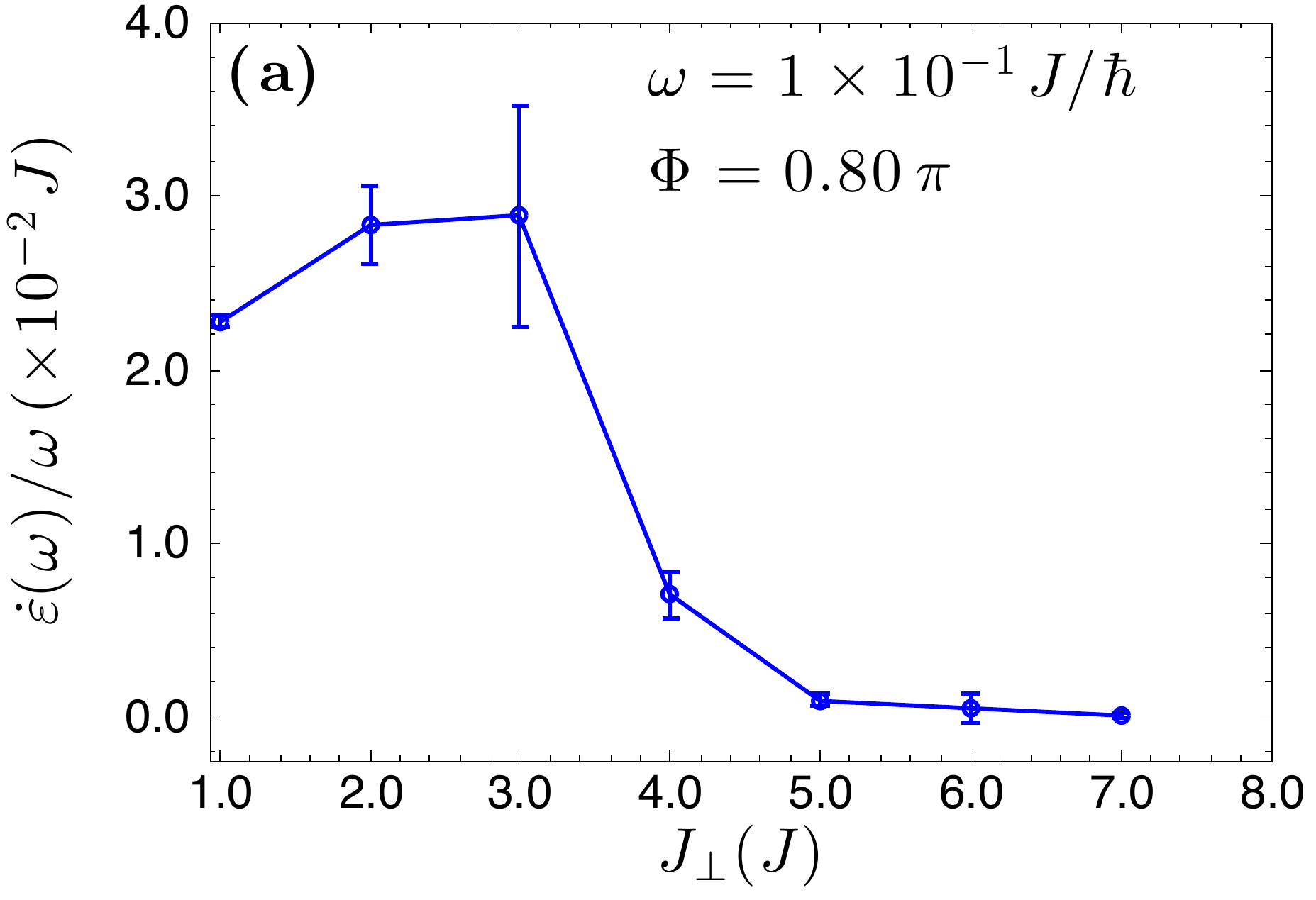}
\includegraphics[width=7.67cm]{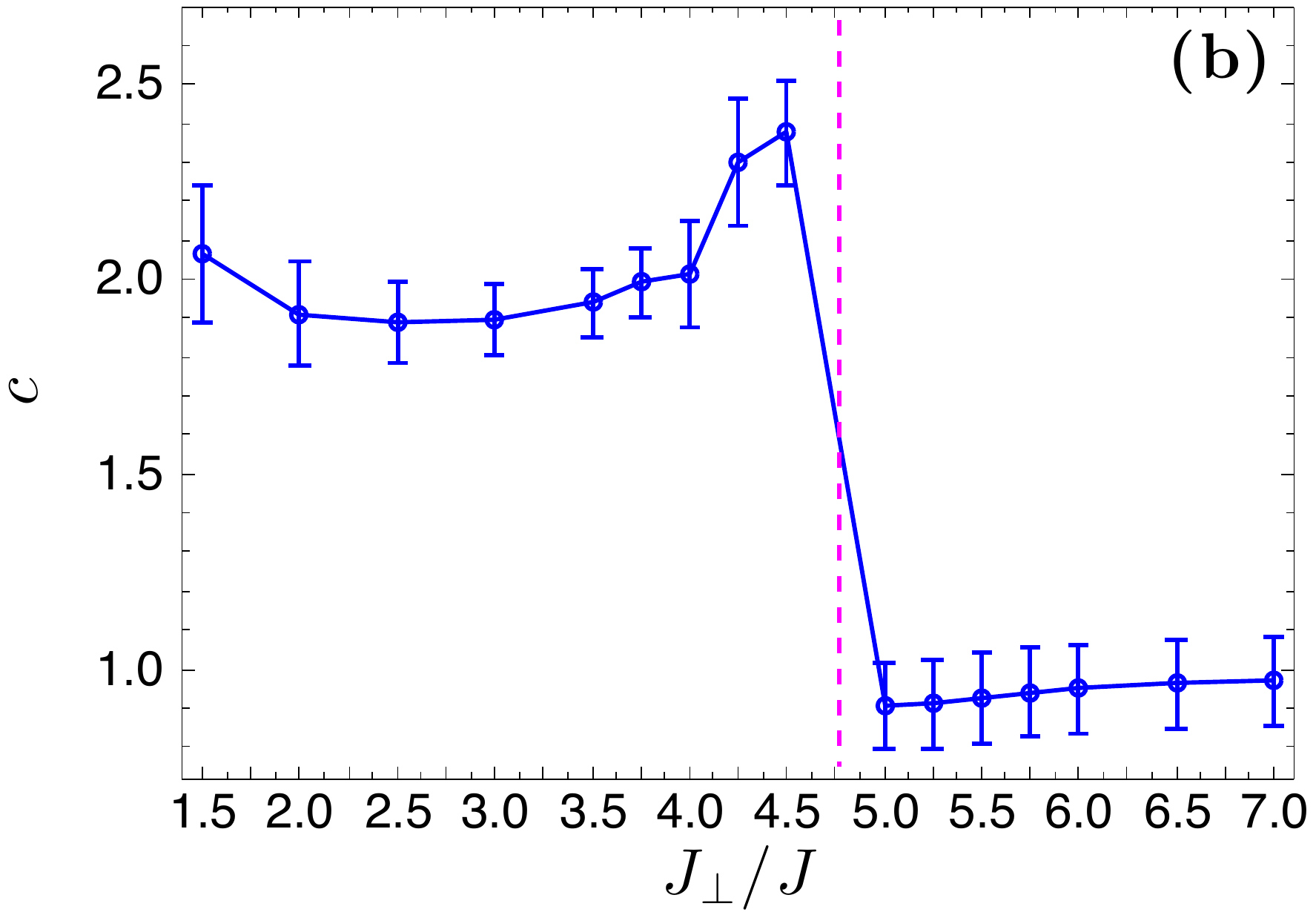}
\caption{{\textbf{(a)} EAR per unit frequency for $n=1/6$ and $L=24$. Other simulations parameters as in Figure~\ref{fig:eardilutecase}\textbf{(e)} and Figure~\ref{fig:earmanybodycase}\textbf{(d)}. Here, the energy absorption starts to be suppressed between $J_{\perp}=4.0\,J$ and $J_{\perp}=5.0\,J$, i.e., at an intermediate value between the critical point at $n=1/12$ ($J_{\perp,c}\simeq 5.9\,J$) and the one at $n=1/4$ ($J_{\perp,c}\simeq 3.8\,J$). \textbf{(b)} Central charge $c$ for hard-core bosons at $n=1/6$ as in panel \textbf{(a)}, but using $L=96$. We see that $c$ drops from $c=2$ (vortex phase) to $c=1$ (Meissner phase) between $J_\perp=4.5\,J$ and $J_\perp=5.0\,J$ (magenta line), in agreement with the value estimated by looking at the EAR per unit frequency in panel \textbf{(a)}. We conclude that the behaviour of the EAR per unit frequency in panel \textbf{(a)} correctly signals the opening of the spin gap also for this value of the density.}}
\label{fig:earccintermediatefillings}
\end{figure}

{The data of the EAR per unit frequency as a function of $J_\perp/J$ are shown in Figure~\ref{fig:earccintermediatefillings}\textbf{(a)}: the energy absorption is nonzero for $J_\perp/J\lesssim4.0$, and it starts to be suppressed between $J_{\perp}\simeq4.0\,J$ and $J_{\perp}\simeq5.0\,J$, suggesting that the spin gap opens between these two value of $J_\perp/J$. As we did for the $n=1/4$, we compare this result with the bahaviour of the central charge (Figure~\ref{fig:earccintermediatefillings}\textbf{(b)}), computed simulating hard-core bosons at $n=1/6$ and $L=96$. As evident from the figure, the central charge drops from values which are close to $c=2$ (vortex phase) to values close to $c=1$ (Meissner phase) between $J_\perp=4.5\,J$ and $J_\perp=5.0\,J$, in agreement with the value estimated by looking at the EAR per unit frequency in panel \textbf{(a)}.} In the light of these results {together with the results discussed in the previous appendices}, we {are confident about the reliability of the computed energy change $\Delta E(t)$ and EAR $\dot\varepsilon(\omega)$}.


\section*{References}
\providecommand{\newblock}{}


\begin{thebibliography}{10}
\expandafter\ifx\csname url\endcsname\relax
  \def\url#1{{\tt #1}}\fi
\expandafter\ifx\csname urlprefix\endcsname\relax\def\urlprefix{URL }\fi
\providecommand{\eprint}[2][]{\url{#2}}

\bibitem{borhcollectedworks}
Bohr N 1972 {\em Niels Bohr Collected Works\/} vol~1 (Elsevier Science)

\bibitem{jphysradium2361}
Van~Leeuwen H~J 1921 {\em J. Phys. Radium\/} {\bf 2} 361--377

\bibitem{landau2013statistical}
Landau L and Lifshitz E 2013 {\em Statistical Physics\/} v. 5 (Elsevier
  Science)

\bibitem{PhysRevLett.45.494}
Klitzing K~v, Dorda G and Pepper M 1980 {\em Phys. Rev. Lett.\/} {\bf 45}(6)
  494--497

\bibitem{PhysRevLett.48.1559}
Tsui D~C, Stormer H~L and Gossard A~C 1982 {\em Phys. Rev. Lett.\/} {\bf
  48}(22) 1559--1562

\bibitem{giamarchi2003quantum}
Giamarchi T 2003 {\em Quantum Physics in One Dimension\/} International Series
  of Monographs on Physics (Clarendon Press)

\bibitem{RevModPhys.77.259}
Schollw\"ock U 2005 {\em Rev. Mod. Phys.\/} {\bf 77}(1) 259--315

\bibitem{Schollwock201196}
Schollw\"ock U 2011 {\em Ann. Phys.\/} {\bf 326} 96 -- 192

\bibitem{NatPhys10.588.14}
Atala M, Aidelsburger M, Lohse M, Barreiro J~T, Paredes B and Bloch I 2014 {\em
  Nat. Phys.\/} {\bf 10} 588

\bibitem{PhysRevB.64.144515}
Orignac E and Giamarchi T 2001 {\em Phys. Rev. B\/} {\bf 64}(14) 144515

\bibitem{PhysRevB.63.180508}
Donohue P and Giamarchi T 2001 {\em Phys. Rev. B\/} {\bf 63} 180508

\bibitem{PhysRevB.72.104521}
Granato E 2005 {\em Phys. Rev. B\/} {\bf 72}(10) 104521

\bibitem{PhysRevB.73.100502}
Rizzi M, Cataudella V and Fazio R 2006 {\em Phys. Rev. B\/} {\bf 73}(10) 100502

\bibitem{PhysRevA.85.041602}
Dhar A, Maji M, Mishra T, Pai R~V, Mukerjee S and Paramekanti A 2012 {\em Phys.
  Rev. A\/} {\bf 85}(4) 041602

\bibitem{PhysRevLett.111.150601}
Petrescu A and Le~Hur K 2013 {\em Phys. Rev. Lett.\/} {\bf 111}(15) 150601

\bibitem{PhysRevB.87.174501}
Dhar A, Mishra T, Maji M, Pai R~V, Mukerjee S and Paramekanti A 2013 {\em Phys.
  Rev. B\/} {\bf 87}(17) 174501

\bibitem{PhysRevA.89.063617}
Wei R and Mueller E~J 2014 {\em Phys. Rev. A\/} {\bf 89}(6) 063617

\bibitem{1367-2630-16-7-073005}
Tokuno A and Georges A 2014 {\em New J. Phys.\/} {\bf 16} 073005

\bibitem{PhysRevB.91.054520}
Petrescu A and Le~Hur K 2015 {\em Phys. Rev. B\/} {\bf 91}(5) 054520

\bibitem{DiDio2015}
Di~Dio M, Citro R, De~Palo S, Orignac E and Chiofalo M~L 2015 {\em The European
  Physical Journal Special Topics\/} {\bf 224} 525--531

\bibitem{PhysRevB.91.140406}
Piraud M, Heidrich-Meisner F, McCulloch I~P, Greschner S, Vekua T and
  Schollw\"ock U 2015 {\em Phys. Rev. B\/} {\bf 91} 140406

\bibitem{PhysRevA.92.013625}
Uchino S and Tokuno A 2015 {\em Phys. Rev. A\/} {\bf 92}(1) 013625

\bibitem{PhysRevB.92.060506}
Di~Dio M, De~Palo S, Orignac E, Citro R and Chiofalo M~L 2015 {\em Phys. Rev.
  B\/} {\bf 92} 060506

\bibitem{PhysRevB.92.115446}
Cornfeld E and Sela E 2015 {\em Phys. Rev. B\/} {\bf 92}(11) 115446

\bibitem{1367-2630-17-9-092001}
Kolley F, Piraud M, McCulloch I~P, Schollw\"ock U and Heidrich-Meisner F 2015
  {\em New J. Phys.\/} {\bf 17} 092001

\bibitem{PhysRevB.92.115120}
Greschner S, Huerga D, Sun G, Poletti D and Santos L 2015 {\em Phys. Rev. B\/}
  {\bf 92}(11) 115120

\bibitem{PhysRevA.92.053623}
Natu S~S 2015 {\em Phys. Rev. A\/} {\bf 92}(5) 053623

\bibitem{PhysRevLett.115.190402}
Greschner S, Piraud M, Heidrich-Meisner F, McCulloch I~P, Schollw\"ock U and
  Vekua T 2015 {\em Phys. Rev. Lett.\/} {\bf 115} 190402

\bibitem{PhysRevA.93.053629}
Uchino S 2016 {\em Phys. Rev. A\/} {\bf 93}(5) 053629

\bibitem{PhysRevA.94.023630}
Bilitewski T and Cooper N~R 2016 {\em Phys. Rev. A\/} {\bf 94}(2) 023630

\bibitem{PhysRevA.94.063628}
Greschner S, Piraud M, Heidrich-Meisner F, McCulloch I~P, Schollw\"ock U and
  Vekua T 2016 {\em Phys. Rev. A\/} {\bf 94} 063628

\bibitem{1367-2630-18-5-055017}
Orignac E, Citro R, Di~Dio M, De~Palo S and Chiofalo M~L 2016 {\em New J.
  Phys.\/} {\bf 18} 055017

\bibitem{PhysRevA.94.063632}
Anisimovas E, Ra\ifmmode \check{c}\else \v{c}\fi{}i\ifmmode~\bar{u}\else
  \={u}\fi{}nas M, Str\"ater C, Eckardt A, Spielman I~B and
  Juzeli\ifmmode~\bar{u}\else \={u}\fi{}nas G 2016 {\em Phys. Rev. A\/} {\bf
  94}(6) 063632

\bibitem{PhysRevX.7.021033}
Calvanese~Strinati M, Cornfeld E, Rossini D, Barbarino S, Dalmonte M, Fazio R,
  Sela E and Mazza L 2017 {\em Phys. Rev. X\/} {\bf 7}(2) 021033

\bibitem{PhysRevB.96.014524}
Petrescu A, Piraud M, Roux G, McCulloch I~P and Le~Hur K 2017 {\em Phys. Rev.
  B\/} {\bf 96}(1) 014524

\bibitem{RevModPhys.83.1523}
Dalibard J, Gerbier F, Juzeli\ifmmode~\bar{u}\else \={u}\fi{}nas G and \"Ohberg
  P 2011 {\em Rev. Mod. Phys.\/} {\bf 83} 1523--1543

\bibitem{0034-4885-77-12-126401}
Goldman N, Juzeli{\={u}}nas G, \"Ohberg P and Spielman I~B 2014 {\em Rep. Prog.
  Phys.\/} {\bf 77} 126401

\bibitem{nature22811}
Tai M~E, Lukin A, Rispoli M, Schittko R, Menke T, Borgnia D, Preiss P~M, Grusdt
  F, Kaufman A~M and Greiner M 2017 {\em Nature\/} {\bf 546} 519 -- 523

\bibitem{science1514}
Stuhl B~K, Lu H~I, Aycock L~M, Genkina D and Spielman I~B 2015 {\em Science\/}
  {\bf 349} 1514--1518

\bibitem{Fangzhao2017a}
An F~A, Meier E~J and Gadway B 2017 {\em Science Advances\/} {\bf 3}, 4

\bibitem{PhysRevLett.112.043001}
Celi A, Massignan P, Ruseckas J, Goldman N, Spielman I~B,
  Juzeli\ifmmode~\bar{u}\else \={u}\fi{}nas G and Lewenstein M 2014 {\em Phys.
  Rev. Lett.\/} {\bf 112}(4) 043001

\bibitem{1367-2630-18-3-035010}
Barbarino S, Taddia L, Rossini D, Mazza L and Fazio R 2016 {\em New J. Phys.\/}
  {\bf 18} 035010

\bibitem{Zhang1467}
Zhang X, Bishof M, Bromley S~L, Kraus C~V, Safronova M~S, Zoller P, Rey A~M and
  Ye J 2014 {\em Science\/} {\bf 345} 1467--1473

\bibitem{NatPhysPagano}
Pagano G, Mancini M, Cappellini G, Lombardi P, Sch\"afer F, Hu H, Liu X~J,
  Catani J, Sias C, Inguscio M and Fallani L 2014 {\em Nat. Phys.\/} {\bf 10}
  198--201

\bibitem{science1510}
Mancini M, Pagano G, Cappellini G, Livi M, Rider M, Catani J, Sias C, Zoller P,
  Inguscio M, Dalmonte M and Fallani L 2015 {\em Science\/} {\bf 349}
  1510--1513

\bibitem{PhysRevLett.117.220401}
Livi L~F, Cappellini G, Diem M, Franchi L, Clivati C, Frittelli M, Levi F,
  Calonico D, Catani J, Inguscio M and Fallani L 2016 {\em Phys. Rev. Lett.\/}
  {\bf 117}(22) 220401

\bibitem{PhysRevB.71.161101}
Narozhny B~N, Carr S~T and Nersesyan A~A 2005 {\em Phys. Rev. B\/} {\bf 71}(16)
  161101

\bibitem{PhysRevB.73.195114}
Carr S~T, Narozhny B~N and Nersesyan A~A 2006 {\em Phys. Rev. B\/} {\bf 73}(19)
  195114

\bibitem{PhysRevB.76.195105}
Roux G, Orignac E, White S~R and Poilblanc D 2007 {\em Phys. Rev. B\/} {\bf
  76}(19) 195105

\bibitem{1367-2630-17-10-105001}
Mazza L, Aidelsburger M, Tu H~H, Goldman N and Burrello M 2015 {\em New J.
  Phys.\/} {\bf 17} 105001

\bibitem{ncomms9134}
Barbarino S, Taddia L, Rossini D, Mazza L and Fazio R 2015 {\em Nat. Comm.\/}
  {\bf 6} 8134

\bibitem{PhysRevB.92.245121}
Budich J~C, Laflamme C, Tschirsich F, Montangero S and Zoller P 2015 {\em Phys.
  Rev. B\/} {\bf 92}(24) 245121

\bibitem{PhysRevA.93.013604}
Lacki M, Pichler H, Sterdyniak A, Lyras A, Lembessis V~E, Al-Dossary O, Budich
  J~C and Zoller P 2016 {\em Phys. Rev. A\/} {\bf 93}(1) 013604

\bibitem{PhysRevA.95.063612}
Ghosh S~K, Greschner S, Yadav U~K, Mishra T, Rizzi M and Shenoy V~B 2017 {\em
  Phys. Rev. A\/} {\bf 95}(6) 063612

\bibitem{arXiv:1707.05715}
Haller A, Rizzi M and Burrello M 2017 {\em arXiv:1707.05715\/}

\bibitem{RevModPhys.80.885}
Bloch I, Dalibard J and Zwerger W 2008 {\em Rev. Mod. Phys.\/} {\bf 80}(3)
  885--964

\bibitem{PhysRevLett.107.255301}
Aidelsburger M, Atala M, Nascimb\`ene S, Trotzky S, Chen Y~A and Bloch I 2011
  {\em Phys. Rev. Lett.\/} {\bf 107}(25) 255301

\bibitem{PhysRevA.74.041604}
Kollath C, Iucci A, McCulloch I~P and Giamarchi T 2006 {\em Phys. Rev. A\/}
  {\bf 74} 041604

\bibitem{PhysRevLett.97.050402}
Kollath C, Iucci A, Giamarchi T, Hofstetter W and Schollw\"ock U 2006 {\em
  Phys. Rev. Lett.\/} {\bf 97} 050402

\bibitem{PhysRevA.73.041608}
Iucci A, Cazalilla M~A, Ho A~F and Giamarchi T 2006 {\em Phys. Rev. A\/} {\bf
  73} 041608

\bibitem{PhysRevLett.97.260401}
Dalla~Torre E~G, Berg E and Altman E 2006 {\em Phys. Rev. Lett.\/} {\bf 97}
  260401

\bibitem{PhysRevB.77.245119}
Berg E, Dalla~Torre E~G, Giamarchi T and Altman E 2008 {\em Phys. Rev. B\/}
  {\bf 77} 245119

\bibitem{10.1088/0953-4075/46/8/085303}
Dalla~Torre E~G 2013 {\em J. Phys. B: At. Mol. Opt. Phys.\/} {\bf 46} 085303

\bibitem{njp.12.033007}
Gerbier F and Dalibard J 2010 {\em New. J. Phys.\/} {\bf 12} 033007

\bibitem{1742-5468-2004-06-P06002}
Calabrese P and Cardy J 2004 {\em J. Stat. Mech.\/} {\bf 2004} P06002

\bibitem{PhysRevB.57.10324}
Arita R, Kuroki K, Aoki H and Fabrizio M 1998 {\em Phys. Rev. B\/} {\bf 57}(17)
  10324--10327

\bibitem{PhysRevB.83.205113}
Moreno A, Muramatsu A and Manmana S~R 2011 {\em Phys. Rev. B\/} {\bf 83}(20)
  205113

\bibitem{PhysRevLett.91.147902}
Vidal G 2003 {\em Phys. Rev. Lett.\/} {\bf 91} 147902

\bibitem{PhysRevLett.93.040502}
Vidal G 2004 {\em Phys. Rev. Lett.\/} {\bf 93} 040502

\bibitem{suzuki.prog.theor.phys.56.1454}
Suzuki M 1976 {\em Prog. Theor. Phys.\/} {\bf 56} 1454

\bibitem{suzuki.j.math.phys.32.400}
Suzuki M 1991 {\em J. Math. Phys.\/} {\bf 32} 400

\bibitem{RevModPhys.87.637}
Ludlow A~D, Boyd M~M, Ye J, Peik E and Schmidt P~O 2015 {\em Rev. Mod. Phys.\/}
  {\bf 87}(2) 637--701

\bibitem{1367-2630-10-7-073015}
Yi W, Daley A~J, Pupillo G and Zoller P 2008 {\em New. J. Phys.\/} {\bf 10}
  073015

\bibitem{foelling2007a}
Folling S, Trotzky S, Cheinet P, Feld M, Saers R, Widera A, Muller T and Bloch
  I 2007 {\em Nature\/} {\bf 448} 1029--1032

\bibitem{ruostekoski2002a}
Ruostekoski J, Dunne G~V and Javanainen J 2002 {\em Phys. Rev. Lett.\/} {\bf
  88} 180401

\bibitem{jaksch2003a}
Jaksch D and Zoller P 2003 {\em New. J. Phys.\/} {\bf 5} 56

\bibitem{goldman2012a}
Goldman N, Beugnon J and Gerbier F 2012 {\em Phys. Rev. Lett.\/} {\bf 108}(25)
  255303

\bibitem{PhysRevB.94.115112}
Lee J, Sachdev S and White S~R 2016 {\em Phys. Rev. B\/} {\bf 94}(11) 115112

\bibitem{PhysRevB.87.115132}
Rodney M, Song H~F, Lee S~S, Le~Hur K and S\o{}rensen E~S 2013 {\em Phys. Rev.
  B\/} {\bf 87}(11) 115132

\bibitem{PhysRevB.92.035154}
Zhuang Y, Changlani H~J, Tubman N~M and Hughes T~L 2015 {\em Phys. Rev. B\/}
  {\bf 92}(3) 035154

\end{thebibliography}
\end{document}